\documentclass{article}

\usepackage{hyperref}

\usepackage{gb4e}
\noautomath

\usepackage{amsmath, braket,cmll}
\usepackage{ amssymb} 
\usepackage{enumerate}
\usepackage{pgf}
\usepackage{tikz}
\usepackage{subcaption}
\usepackage{float}
\floatstyle{boxed}
\restylefloat{figure}
\usepackage{placeins}

\def\id{\mbox{id}}

\DeclareMathSymbol{\mlq}{\mathord}{operators}{``}
\DeclareMathSymbol{\mrq}{\mathord}{operators}{`'}

\newtheorem{thm}{Theorem}

\date{}

\begin{document}
 \title{Cobordisms and  commutative categorial grammars
}
\author{Sergey Slavnov
\\  National Research University Higher School of Economics
\\ sslavnov@yandex.ru\\} \maketitle

	\begin{abstract}
We propose a concrete   surface representation of  abstract categorial grammars in the  category of {\it word cobordisms} or {\it cowordisms} for short, which are certain bipartite graphs decorated with words in a given alphabet, generalizing linear logic proof-nets. We also introduce and study {\it linear logic grammars},  directly based on cobordisms and using classical multiplicative linear logic as a typing system.
	\end{abstract}

\section{Introduction}
The best known categorial grammars are based on noncommutative variants of linear logic, most notably, on Lambek calculus \cite{Lambek} and its variations/extensions.
On the other hand, such formalisms as
{\it abstract categorial grammars} (ACG) \cite{deGroote}, also known  (with minor variations) as {\it $\lambda$-grammars}  \cite{Muskens} or {\it linear grammars}  \cite{PollardMichalicek}, arise from an alternative or, rather, complementary approach, and use ordinary implicational linear logic and linear $\lambda$-calculus. These  can be called ``commutative''  in contrast to the ``noncommutative'' Lambek grammars. Commutative grammars are attractive because of the much more familiar and intuitive  underlying logic, besides they are remarkably expressive.
Unfortunately,   basic constituents of ACG used for syntax generation seem extremely abstract: they are just linear $\lambda$-terms. Identifying an abstract $\lambda$-term  with some element of language is not so easy, and syntactic analysis becomes complicated. It seems that some more concrete {\it surface representation} for commutative grammars would be highly desirable.

In this work we propose that such a representation is indeed possible. We introduce a specific structure of {\it word cobordisms}, or, simply {\it cowordisms}, as we abbreviate  for a joke.

{\it Word cobordism} is a bipartite graph, more precisely, a perfect matching (generalizing linear logic {\it proof-nets}), whose edges are labeled with words in a given alphabet, and whose vertex set is subdivided into the input and the output parts. This can be seen as a one-dimensional topological cobordism (see \cite{Stong_cobordisms}, \cite{BaezTQFT}) decorated with words, which explains our terminology. (For a pedestrian discussion of cobordisms that might be relevant to the content of this paper,  see \cite{BaezRosetta}.)

Just as topological cobordisms, word cobordisms can be organized into a {\it category}, with composition given by gluing inputs to outputs. The resulting category has a rich structure, in particular it is {\it compact closed} (see \cite{KellyLaplaza}, also \cite{AbramskyCoecke}), and, as any compact closed category, it provides a  denotational model for multiplicative linear logic and for linear $\lambda$-calculus.
 The latter model gives rise to   the geometric {\it cowordism representation} of string ACG that we discuss.

On the other hand, the very structure of cowordism category  with its involutive duality, suggests using {\it classical} (rather than intuitionistic) multiplicative linear logic ({\bf MLL}) as more natural for this setting. Thus, we also define and study  {\it linear logic grammars} ({LLG}),  based directly on the cowordism representation and using {\bf MLL} as the typing system. String ACG can be seen as a particular case of LLG.

LLG with their underlying compact category could  be seen  as a  commutative version of {\it pregroup grammars} (see \cite{Lambek_pregroups}).
This suggests possible connections with {\it categorical compositional distributional semantics} ({DisCoCat}) \cite{CoeckeSadrzadehClark}, which use pregroup grammars a lot.
Indeed, {DisCoCat} models are based on  finite-dimensional vector spaces and use their symmetric and compact closed categorical structure in an essential way. Arguably,  LLG   match  these structures just better than other syntactic formalisms.
Although such a matching is not required for any known construction, and we cannot say even if it is useful at all, it seems  at least interesting that a parallel symmetric compact structure  can be found on the syntactic side as well.

(We should add though that the above parallelism does not go to the extreme. Typically, {DisCoCat} models, apart from using the canonical symmetric compact structure of vector spaces, impose additional, non-canonical structure of {\it commutative Frobenius algebra}, so called ``spiders'' \cite{Frobenius_anatomy_I}. This latter commutativity has no analogue on the syntactic, surface level.)

 In any case, we think that the cowordism representation with its simple geometric meaning and diagrammatic reasoning might be helpful for studying language generation, some examples are given. Hopefully, it can also be used for applying ideas of DisCoCat to various ``commutative'' formalisms, thus going beyond context-free languages.

\section{Boundaries and multiwords}
Let $T$ be a finite alphabet.
We denote the set of all finite words in $T$ as $T^*$, and the empty word as $\epsilon$.
For consistency of definitions, we will also need
 {\it cyclic words}, which are equivalence classes of elements of $T^*$ quotiented by cyclic permutations of letters.
For $w\in T^*$  we denote the corresponding cyclic word  as $[w]$.

For a set $X$ of natural numbers and an integer $n$ we use the notation
$n+X=\{n+m|~m\in X\}$, and $n-X=\{n-m|~m\in X\}$. For multisets $A$, $B$ we denote their disjoint union as $A+B$.
For a positive integer $N$, we denote  ${\bf I}(N)=\{1,\ldots,N\}$.

Finally, for positive integers $N,M$ with $M\leq N$, we will  use the {\it shifted embedding} function
 $\mbox{shift}_{k}^{s}:{\bf{I}}(M)\to{\bf I}(N)$, where $M+s\leq N$, defined as
$\mbox{shift}_{k}^{s}(i)=i,\mbox{ if }i<k,~\mbox{shift}_{k}^{s}(i)=i+s\mbox{ if }i\geq k$.

A {\it boundary} $X$ consists of a natural number $|X|$, the {\it cardinality} of $X$, and a subset $X_l\subseteq{\bf I}(|X|)$ of {\it left endpoints} of $X$.
We will  denote ${\bf I}(|X|)={\bf I}(X)$.

The complement  of $X_l$ in ${\bf I}(X)$ is denoted as $X_r$.
Elements of $X_l$ are called {\it left endpoints} of $X$ and are said to have {\it left polarity}, while elements of $X_r$ are {\it right endpoints} of $X$ and  have {\it right polarity}.

A boundary $X$ is, basically, a linearly ordered finite set of cardinality $|X|$, equipped with a partition into left and right endpoints.

For two boundaries $X,Y$ and an integer $i$ such that $i+|Y|\leq|X|$,
we say that
$i+Y$ is a {\it subboundary} of $X$ if $i+Y_l\subseteq X_l$ and $i+Y_r\subseteq X_r$.

Given two boundaries $X$ and $Y$, the {\it tensor product boundary} $X\otimes Y$  and the {\it dual boundary} $X^\bot$ are obtained, respectively, by concatenation and order and polarity reversal,
i.e.
$|{X\otimes Y}|=|X|+|Y|$, $(X\otimes Y)_l=X_l\cup(|X|+Y_l)$, and
 $|{X^\bot}|=|X|$, $(X^\bot)_l=|X|+1-X_r$.
 The neutral element for tensor product  is the {\it unit boundary} ${\bf 1}$  defined by
$|{\bf 1}|=0$, ${\bf 1}_l=\emptyset$.

Note that we have the identity
$(X\otimes Y)^\bot=Y^\bot\otimes X^\bot$.
This should not suggest any sort of noncommutativity in the category of boundaries. We will have a natural isomorphism between $(X\otimes Y)^\bot$ and $X^\bot\otimes Y^\bot$, just not equality. The flip of tensor factors will allow somewhat better pictures, with fewer crossings.

Given an alphabet $T$, a {\it regular multiword} $M$ over   $T$  with boundary $X$ is a  directed   graph on the set ${\bf I}(X)$ of vertices, whose  edges are labelled with words in $T^*$,   such that each vertex is adjacent
to exactly one edge (so that it is a perfect matching), and for every edge its left endpoint is in $X_l$ and its right endpoint is in $X_r$.

In the following we will identify a regular multiword with the set of its labeled edges. The notation $[x,w,y]$ will stand for an edge from $x$ to $y$ labeled with the word $w$.

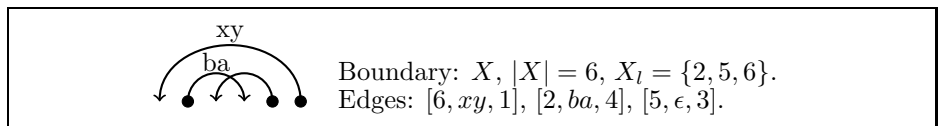
\begin{figure}
\centering
         \begin{tikzpicture}[xscale=1.5,yscale=1.5]
    \begin{scope}
        \draw[thick,->](1.25,0) to  [out=90,in=0] (.625,.5) to  [out=180,in=90](0,0);
        \node at(.625,.6){xy};
        \node at(.5,.35){ba};
        \draw[thick,->](1,0) to  [out=90,in=0] (.75,.25) to  [out=180,in=90](0.5,0);
        \draw[thick,<-](0.75,0) to  [out=90,in=0] (.5,.25) to  [out=180,in=90](.25,0);
        \draw [fill] (.25,0) circle [radius=0.05];
        \draw [fill] (1,0) circle [radius=0.05];
        \draw [fill] (1.25,0) circle [radius=0.05];
        \end{scope}
        \node[right] at (1.5,.25){Boundary: $X$, $|X|=6$, $X_l=\{2,5,6\}$.};
         \node[right] at (1.5,.0) {Edges:  $[6,xy,1]$, $[2,ba,4]$, $[5,\epsilon,3]$.};

        \end{tikzpicture}
        \caption{Multiword}
        \label{multiword_example}
   \end{figure}
A general {\it multiword} $M$ over  $T$ with boundary $X$  is defined as a pair $M=(M_0,M_c)$, where $M_0$, the {\it regular} part, is a regular multiword over $T$ with the boundary $X$, and $M_c$, the {\it singular} part, is a finite multiset of cyclic words over $T$. A multiword is {\it acyclic} or {\it regular} if its singular part is empty. Otherwise it is {\it singular}.

Singular multiwords should be understood as pathological (in the context of this work), but we need them for consistency of definitions.
Geometrically, a multiword can be understood as the disjoint union of an edge-labeled graph and a collection of closed curves (i.e. circles) labeled with cyclic words.

  We will use certain conventions for depicting  multiwords, which guarantee unambiguous reading of pictures. Unless otherwise stated,  points of the boundary are ordered from left to right. Left endpoints  are marked as solid dots, and right endpoints as arrowheads. Also, our strict convention for reading edge labels is that {\it words in a picture are always read from left to right}, in the usual way, no matter what is the direction of edges.
 An example is in Figure \ref{multiword_example}.

 Given two multiwords $M=(M_0,M_c)$ and $N=(N_0,N_c)$  with boundaries $X$ and $Y$ respectively, the {\it tensor product  multiword} $M\otimes N$ has boundary $X\otimes Y$ and is defined as the disjoint union, i.e.
  $(M\otimes N)_c=M_c+ N_c$
   and
 $(M\otimes N)_0=\{[i,w,j]|~ [i,w,j]\in M_0\}\cup\{[|X|+i,w,|X|+j]|~ [i,w,j]\in N_0\}$.

A crucial operation on multiwords is {\it  contraction}, which consists in gluing neighboring endpoints of opposite polarity and concatenating the corresponding edge labels in the direction from the left endpoint to the right. Here is an accurate definition.

Let $M$ be a multiword  with  boundary $X$, and $n<|X|$   be such that $n$ and
$n+1$ have opposite polarity in $X$. Let $x$ be the right endpoint in the pair $(n,n+1)$  and  $y$ be the left one.

The {\it elementary contraction} $\langle M\rangle_{n,n+1}$ of $M$ along  $n$ and $n+1$ is the
multiword $M'$ with the boundary $X'$, where $|{X'}|=|X|-2$, $(X')_l=(\mbox{shift}_{n}^{2})^{-1}(X_l)$,  constructed as follows.

If $x$ and $y$ are not connected with an edge in $M_0$, then
$M_c'=M_c$, and $M'_0$ consists of all  edges
$[i,w,j]$ such that $[\mbox{shift}_{n}^{2}(i),w,\mbox{shift}_{n}^{2}(j)]\in M_0$ plus the edge
$[\alpha,uv,\beta]$ such that $[\mbox{shift}_{n}^{2}(\alpha),u,x]$, $[y,v,\mbox{shift}_{n}^{2}(\beta)]$ are in $M_0$.

If there is an edge $[y,w,x]\in M_0$, then
$M_c'=M_c+\{[w]\}$,
and $M'_0$ consists of all  edges
$[i,w,j]$ such that $[\mbox{shift}_{n}^{2}(i),w,\mbox{shift}_{n}^{2}(j)]\in M_0$.

It is easy to see that in all cases $M_0'$ is a perfect matching and its edges start at left endpoints of $X_l'$.
Also, when the contracted vertices $x$ and $y$ happen to be  connected with an edge, the resulting multiword necessarily is  singular.

Elementary contractions can be iterated.

Let $X$, $Y$ be  boundaries, $i\in {\bf N}$ and assume that $i+Y^\bot\otimes Y$ is a subboundary of $X$.
Let $n=|Y|=|{Y^\bot}|$.
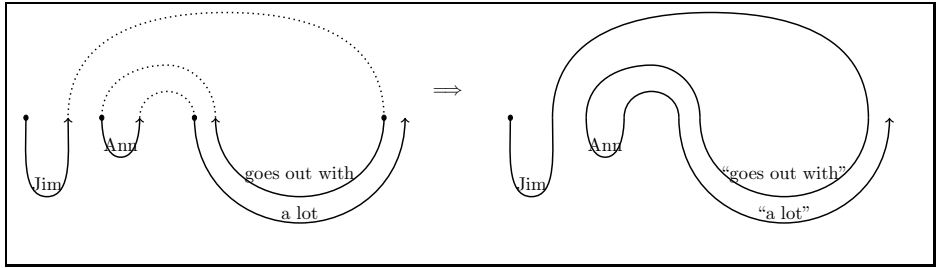
\begin{figure}
\scalebox{.7}
{
     \begin{tikzpicture}[xscale=.8]

        \draw[thick,dotted,-](8,0) to  [out=90,in=0] (6.85,1) to  [out=180,in=90](5.3,0);
        \draw[thick,->](5.3,0) to  [out=-90,in=180] (5.75,-0.75) to  [out=0,in=-90] (6.2,0);
        \node[above] at (5.75,-0.75) {$\mbox{Ann}$};
        \draw[thick,dotted,-](6.2,0)to  [out=90,in=180] (6.85,0.5) to  [out=0,in=90] (7.5,0);
        \draw [fill] (5.3,0) circle [radius=0.05];
        \node[above] at (4,-1.5) {$\mbox{Jim}$};

         \draw[thick,->](3.5,0) to  [out=-90,in=180] (4,-1.5) to  [out=0,in=-90] (4.5,0);
         \draw [fill] (3.5,0) circle [radius=0.05];

        \draw[thick,dotted,-]
        (12,0) to  [out=90,in=0] (8.25,2) to  [out=180,in=90](4.5,0);


        \begin{scope}[shift={(1,-0.75)}]
            \draw[thick,->](6.5,.75) to  [out=-90,in=180] (9,-1.25) to  [out=0,in=-90] (11.5,.75);
            \draw[thick,<-](7,.75) to  [out=-90,in=180] (9,-.75) to  [out=0,in=-90] (11,.75);
            \draw [fill] (6.5,.75) circle [radius=0.05];
            \draw [fill] (11,.75) circle [radius=0.05];
            \node[above] at (9,-.6) {$\mbox{goes out with}$};
            \node[above] at (9,-1.3) {$\mbox{a lot}$};
            \node [right] at (12,1.25) {$\Longrightarrow$};
        \end{scope}

        \begin{scope}[shift={(11.5,0)}]
        \draw[thick,-](8,0) to  [out=90,in=0] (6.85,1) to  [out=180,in=90](5.3,0);
        \draw[thick,-](5.3,0) to  [out=-90,in=180] (5.75,-0.75) to  [out=0,in=-90] (6.2,0);
        \node[above] at (5.75,-0.75) {$\mbox{Ann}$};
        \draw[thick,-](6.2,0)to  [out=90,in=180] (6.85,0.5) to  [out=0,in=90] (7.5,0);
        \node[above] at (4,-1.5) {$\mbox{Jim}$};

         \draw[thick,-](3.5,0) to  [out=-90,in=180] (4,-1.5) to  [out=0,in=-90] (4.5,0);
         \draw [fill] (3.5,0) circle [radius=0.05];

        \draw[thick,-]
        (12,0) to  [out=90,in=0] (8.25,2) to  [out=180,in=90](4.5,0);


        \begin{scope}[shift={(1,-0.75)}]
            \draw[thick,->](6.5,.75) to  [out=-90,in=180] (9,-1.25) to  [out=0,in=-90] (11.5,.75);
            \draw[thick,-](7,.75) to  [out=-90,in=180] (9,-.75) to  [out=0,in=-90] (11,.75);
            \node[above] at (9,-.6) {$\mbox{``goes out with''}$};
            \node[above] at (9,-1.3) {$\mbox{``a lot''}$};
        \end{scope}

        \end{scope}

    \end{tikzpicture}
    }\\
    \caption{Iterated contractions}
\label{iterated_contraction}
\end{figure}
Then for any multiword $M$ with boundary $X$  we define the {\it iterated contraction} $\langle M\rangle_{i+Y^\bot\otimes Y}$ of $M$ along ${i+Y^\bot\otimes Y}$ by
$\langle M\rangle_{i+Y^\bot\otimes Y}=\langle\ldots\langle~ \langle~\langle M\rangle_{i+n,i+n+1}\rangle_{i+n-1,i+n}\rangle\ldots\rangle_{i+1,i+2}$.

It is easy to check that the above is well defined.

In order to avoid  possible ambiguity in pictures with iterated contractions,  we will use quotation  marks.
An example  is shown in Figure \ref{iterated_contraction},
 where dotted lines connect neighboring vertices that will be contracted.
 When we replace dotted lines with solid ones, the resulting graph has discontinuous edge labels, and it is not immediately clear how to read them.
  Our convention is that any block in quotation marks is read from left to right, as usual, while several blocks labeling one edge are read in the order in which they appear as we traverse the edge from the left endpoint to the right one.
 In particular, in Figure \ref{iterated_contraction}, when all the zigzagging is reduced, we obtain the  sentence
 ${\mbox{``Jim goes out with Ann a lot''}}$.

\section{Word cobordisms}
Let $X,Y$ be boundaries and $T$ be an alphabet.

A {\it word cobordism} or, simply, a {\it cowordism} $\sigma:X\to Y$
 over  $T$ from $X$ to $Y$ is  a multiword over $T$ with  boundary $X^\bot\otimes Y$.
We say that $Y$ is the {\it outgoing boundary} of $\sigma$, and $X$ is the {\it incoming   boundary}.
A cowordism is {\it regular} if its underlying multiword is regular,  otherwise it is {\it singular}.

When depicting a cowordism $\sigma:X\to Y$,
we put elements $1,\ldots,|X|$ of the  subboundary $X^\bot$ on one vertical line, with the  increasing order  corresponding to the direction {\it up}, and
we put the elements $|X|+1,\ldots,|X|+|Y|$ of the  subboundary $|X|+Y$ on a parallel line to the right, in the increasing order corresponding to the direction {\it down}.

\begin{figure}
\begin{subfigure}{.4\textwidth}
\centering
\begin{tikzpicture}
\begin{scope}[shift={(-.7,0)}]
\draw[dashed](0.25,1.2)--(0,1.2)--(0,.3)--(0.25,.3);
\node[left]at(0,.75){$X$};
\end{scope}
\draw[thick,->](-.05,0.3)--(0.25,0.3);
\draw[thick,<-](-.05,0.6)--(0.25,0.6);
\draw[thick,->](-.05,.9)--(0.25,.9);
\draw[thick,->](-.05,1.2)--(0.25,1.2);

\node[left] at(-.05,0.3){$1$};
\node[left] at(-.05,0.6){$2$};
\node[left] at(-.05,0.9){$3$};
\node[left] at(-.05,1.2){$4$};

\node[right] at(2.3,0.3){$8$};
\node[right] at(2.3,0.6){$7$};
\node[right] at(2.3,0.9){$6$};
\node[right] at(2.3,1.3){$5$};

\draw [fill] (2.3,.9) circle [radius=0.05];
\draw [fill] (-.05,0.3) circle [radius=0.05];
\draw [fill] (-.050,.9) circle [radius=0.05];
\draw [fill] (-.05,1.2) circle [radius=0.05];

\draw[draw=black,fill=gray!10](0.25,0)rectangle(2,1.5);\node at(1.125,.75){$\sigma$};
\draw[thick,->](2,0.3)--(2.3,0.3);
\draw[thick,->](2,.6)--(2.3,0.6);
\draw[thick,<-](2,.9)--(2.3,.9);
\draw[thick,->](2,1.2)--(2.3,1.2);
\begin{scope}[shift={(2.7,0)}]
\draw[dashed](0.,1.2)--(0.25,1.2)--(0.25,.3)--(0.,.3);
\node[right]at(0.25,.75){$Y$};
\end{scope}
\end{tikzpicture}
\caption{Detailed picture}
\label{cowordism}
\end{subfigure}
\begin{subfigure}{.4\textwidth}
\centering
\begin{tikzpicture}[scale=.8]
\node[left] at(-1.25,-0.){$\sigma:$};
\draw[thick](0,.25)--(-0.5,.25);

\node[left] at(-0.5,.25){$X_1$};

\draw[thick](1,.25)--(1.5,.25);
\node[right] at(1.5,0.25){$Y_1$};

\node[right] at(1,0){$\cdots$};
\node[left] at(0,0){$\cdots$};

\draw[draw=black,fill=gray!10](0.,-.5)rectangle(1,.5);\node at(.5,0){$\tau$};

\begin{scope}[shift={(0,-.5)}]

\draw[thick](0,.25)--(-0.5,.25);

\node[left] at(-0.5,.25){$X_n$};

\draw[thick](1,.25)--(1.5,.25);
\node[right] at(1.5,0.25){$Y_m$};

\end{scope}
\end{tikzpicture}
\caption{Many inputs and outputs}
\label{many_inputs}
\end{subfigure}\\

\begin{subfigure}{1\textwidth}
\begin{tikzpicture}[scale=.7]
\node[left] at(-1,.25){$\sigma:$};
\draw[thick](0,.25)--(-0.5,.25);

\node[left] at(-0.5,.25){$X$};

\draw[thick](1,.25)--(1.5,.25);
\node[right] at(1.5,0.25){$Y$};

\draw[draw=black,fill=gray!10](0.,0)rectangle(1,.5);\node at(.5,.25){$\sigma$};

\begin{scope}[shift={(4.5,0)}]
\node[left] at(-1,.25){$\tau:$};
\draw[thick](0,.25)--(-0.5,.25);

\node[left] at(-0.5,.25){$Y$};

\draw[thick](1,.25)--(1.5,.25);
\node[right] at(1.5,0.25){$Z$};

\draw[draw=black,fill=gray!10](0.,0)rectangle(1,.5);\node at(.5,.25){$\tau$};
\end{scope}

\begin{scope}[shift={(1.5,0)}]
\begin{scope}[shift={(8.5,0)}]
\node[left] at(-1,.25){$\tau\circ\sigma:$};
\draw[thick](0,.25)--(-0.5,.25);

\node[left] at(-0.5,.25){$X$};

\draw[thick](1,.25)--(1.5,.25);

\draw[draw=black,fill=gray!10](0.,0)rectangle(1,.5);\node at(.5,.25){$\sigma$};

\begin{scope}[shift={(1.5,0)}]

\draw[thick](1,.25)--(1.5,.25);
\node[right] at(1.5,0.25){$Z$};

\draw[draw=black,fill=gray!10](0.,0)rectangle(1,.5);\node at(.5,.25){$\tau$};
\end{scope}
\end{scope}

\end{scope}
\end{tikzpicture}
\caption{Composition, schematic picture}
\label{composition}
\end{subfigure}\\

\begin{subfigure}{1\textwidth}
\scalebox{.9}
{
\begin{tikzpicture}[scale=.5]
\node[left]at(-.7,.75){$\sigma:$};
\draw[dashed](0.25,1.2)--(0,1.2)--(0,.3)--(0.25,.3);
\node[left]at(0,.75){$X$};
    \begin{scope}[shift={(.8,0)}]
    \draw[thick,->](-.3,0.3)--(0.25,0.3);
    \draw[thick,<-](-.3,0.6)--(0.25,0.6);
    \draw[thick,->](-.3,.9)--(0.25,.9);
    \draw[thick,->](-.3,1.2)--(0.25,1.2);

    %

    \draw [fill] (2.55,.9) circle [radius=0.05];
    \draw [fill] (-.3,0.3) circle [radius=0.05];
    \draw [fill] (-.30,.9) circle [radius=0.05];
    \draw [fill] (-.3,1.2) circle [radius=0.05];

    \draw[draw=black,fill=gray!10](0.25,0)rectangle(2,1.5);\node at(1.125,.75){$\sigma$};
    \draw[thick,->](2,0.3)--(2.55,0.3);
    \draw[thick,->](2,.6)--(2.55,0.6);
    \draw[thick,<-](2,.9)--(2.55,.9);
    \draw[thick,->](2,1.2)--(2.55,1.2);
    \end{scope}

\begin{scope}[shift={(3.5,0)}]
\draw[dashed](0.,1.2)--(0.25,1.2)--(0.25,.3)--(0.,.3);
\node[right]at(0.25,.75){$Y$};
\end{scope}

    \begin{scope}[shift={(7,0)}]
    \node[left]at(-.7,.75){$\tau:$};
    \draw[dashed](0.25,1.2)--(0,1.2)--(0,.3)--(0.25,.3);
    \node[left]at(0,.75){$Y$};
        \begin{scope}[shift={(.8,0)}]
        \draw[thick,->](-.3,0.3)--(0.25,0.3);
        \draw[thick,->](-.3,0.6)--(0.25,0.6);
        \draw[thick,<-](-.3,.9)--(0.25,.9);
        \draw[thick,->](-.3,1.2)--(0.25,1.2);


        \node[right] at(2.4,.75){$Z$}; 

        \draw [fill] (-.3,0.3) circle [radius=0.05];
        \draw [fill] (-.3,.6) circle [radius=0.05];
        \draw [fill] (-.3,1.2) circle [radius=0.05];

        \draw[draw=black,fill=gray!10](0.25,0)rectangle(2,1.5);\node at(1.125,.75){$\tau$};
        \draw[thick,->](2.,0.6)--(2.55,0.6);
        \draw[thick,<-](2.,.9)--(2.55,.9);
        \draw [fill] (2.55,.9) circle [radius=0.05];


        \end{scope}

        \end{scope}
        \begin{scope}[shift={(15,0)}]
        \node[left]at(-.8,.75){$\tau\circ\sigma:$};
        \draw[dashed](0.25,1.2)--(0,1.2)--(0,.3)--(0.25,.3);
        \node[left]at(0,.75){$X$};
        \begin{scope}[shift={(.8,0)}]
        \draw[thick,->](-.3,0.3)--(0.25,0.3);
        \draw[thick,<-](-.3,0.6)--(0.25,0.6);
        \draw[thick,->](-.3,.9)--(0.25,.9);
        \draw[thick,->](-.3,1.2)--(0.25,1.2);

        %

        \draw [fill] (-.3,0.3) circle [radius=0.05];
        \draw [fill] (-.3,.9) circle [radius=0.05];
        \draw [fill] (-.3,1.2) circle [radius=0.05];

        \draw[draw=black,fill=gray!10](0.25,0)rectangle(2,1.5);\node at(1.125,.75){$\sigma$};
        \draw[thick,-](2,0.3)--(3.5,0.3);
        \draw[thick,-](2,.6)--(3.5,0.6);
        \draw[thick,-](2,.9)--(3.5,.9);
        \draw[thick,-](2,1.2)--(3.5,1.2);
        \end{scope}

        \begin{scope}[shift={(3,0)}]

            \begin{scope}[shift={(.8,0)}]
            \draw[thick](-.05,0.3)--(0.25,0.3);
            \draw[thick,-](-.05,0.6)--(0.25,0.6);
            \draw[thick,-](-.05,.9)--(0.25,.9);
            \draw[thick](-.05,1.2)--(0.25,1.2);

            \node[right] at(2.4,.75){$Z$};

            \draw[draw=black,fill=gray!10](0.25,0)rectangle(2,1.5);\node at(1.125,.75){$\tau$};

            \draw[thick,->](2.,0.6)--(2.55,0.6);
            \draw[thick,<-](2.,.9)--(2.55,.9);
            \draw [fill] (2.55,.9) circle [radius=0.05];

            \end{scope}

        \end{scope}

    \end{scope}
\end{tikzpicture}
}
\caption{Composition, detailed picture}
\label{composition_detailed}
\end{subfigure}
\caption{Cowordisms}
\end{figure}
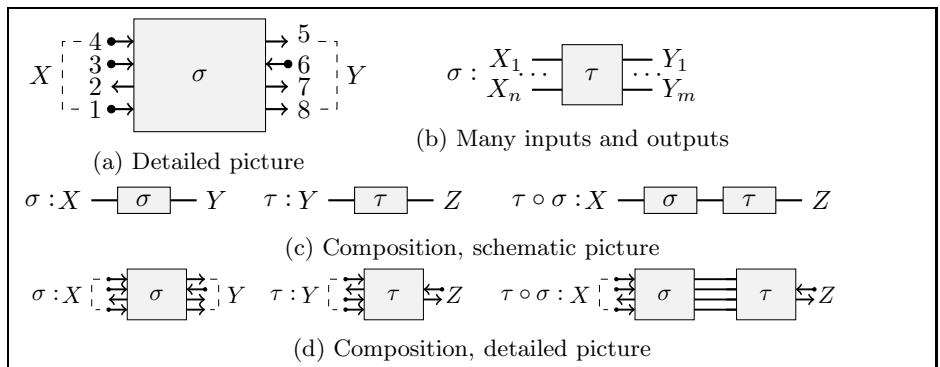

For example, if the boundaries $X,Y$ are given by
\begin{equation}\label{graph.lang.example}
|X|=4,\quad X_l=\{3\},\quad |Y|=4,\quad Y_l=\{2\},
\end{equation}
then a cowordism $\sigma:X\to Y$ will be depicted as in Figure \ref{cowordism}  (where we indicate vertex numbers for clarity). The subboundary $X^\bot$ of $\sigma$ corresponds to the incoming boundary $X$ by means of an order and polarity reversing bijection. In particular the {\it right} endpoint $2$ in the picture corresponds to the {\it left} endpoint $3$ of $X$.

In general, when the structure of boundaries is not important, we ``squeeze'' parallel edges into one and represent a cowordism $\sigma:X\to Y$ schematically as a box with an incoming wire labeled with $X$ and an outgoing wire labeled with $Y$. More generally,
we represent a cowordism
$\sigma:X_1\otimes\ldots\otimes X_n\to Y_1\otimes\ldots\otimes Y_m$
 as a box whose $n$ incoming wires are labeled with $X_i$'s and $m$ outgoing wires are labeled with $Y_i$'s, as in Figure \ref{many_inputs}.
Such a ``squeezed'' picture is consistent with the full picture. If we ``expand'' each edge into  parallel edges adjacent to  points in the corresponding subboundary, we obtain the detailed picture.
When we depict a cowordism $\sigma:{\bf 1}\to X$, respectively  $\tau:X\to{\bf 1}$, we do not have wires on the left, respectively right.

This is, of course, a variation of the familiar {\it pictorial} language for monoidal categories. Note, however, that, since cowordisms are, by definition, geometric objects, the diagrammatic representation is quite {\it literal}, and diagrammatic reasoning is valid automatically, without further justification.

Matching cowordisms are composed  by gluing incoming and outgoing boundaries.

Let boundaries $X,Y,Z$ and cowordisms
$\sigma:X\to Y$, $\tau:Y\to Z$, with the underlying multiwords $M_\sigma$, $M_\tau$ respectively be given.
The {\it composition} $\tau\circ\sigma:X\to Z$
is the cowordism
whose underlying multiword $M_{\tau\circ\sigma}$ is obtained as the iterated contraction
$M_{\tau\circ\sigma}=\langle M_\sigma\otimes M_\tau\rangle_{|X|+Y\otimes Y^\bot}$.

It is easy to see that, with our conventions, composition of cowordisms $\sigma:X\to Y$, $\tau:Y\to Z$ corresponds to the  schematic picture in Figure \ref{composition}.

We get a detailed, ``full'' picture by expanding every edge into as many  parallel edges as there are points in the corresponding boundary.
For example, if $X,Y$ are as in (\ref{graph.lang.example}), and $Z$, say, has two points of opposite polarity, then the schematic picture in Figure \ref{composition} translates to the detailed picture in Figure \ref{composition_detailed}.

It is evident from geometric representation that composition of cowordisms is associative.

The {\it identity cowordism} $\id_X:X\to X$
is  the regular multiword with the boundary $X^\bot\otimes X$ defined as
$$\id_X=\{[|X|+i,\epsilon,|X|-i+1]|~i\in X_l\}\cup\{[|X|-i+1,\epsilon,|X|+i]|~i\in X_r\}.$$

In a schematic, ``squeezed'' picture, the identity cowordism corresponds to a single wire:
\begin{tikzpicture}[baseline=8pt,scale=.5]
\node[left]at(0.,.75){$\id_X:$};
\node[left]at(0.75,.75){$X$};
\draw[thick](.75,0.75)--(1.75,0.725);
\node[right]at(1.75,.75){$X$};
\end{tikzpicture}.
In the full picture there are as many parallel wires as there are points in $X$.
If $X$ is as in (\ref{graph.lang.example}), then the  full picture is the following:
\begin{tikzpicture}[baseline=8pt,scale=.5]
\begin{scope}[shift={(-7.55,0)}]
\node[left]at(-.7,.75){$\id_X:$};
\draw[dashed](0.1,1.2)--(0,1.2)--(0,.3)--(0.1,.3);
\node[left]at(0,.75){$X$};
\begin{scope}[shift={(.4,0)}]
\draw[thick,->](-.05,0.3)--(1.45,0.3);
\draw[thick,<-](-.05,0.6)--(1.45,0.6);
\draw[thick,->](-.05,.9)--(1.45,.9);
\draw[thick,->](-.05,1.2)--(1.45,1.2);

\draw [fill] (1.45,.6) circle [radius=0.05];
\draw [fill] (-.05,0.3) circle [radius=0.05];
\draw [fill] (-.050,.9) circle [radius=0.05];
\draw [fill] (-.05,1.2) circle [radius=0.05];
\draw[dashed](1.55,1.2)--(1.7,1.2)--(1.7,.3)--(1.55,.3);
\node[right]at(1.7,.75){$X$};
\end{scope}
\end{scope}
\end{tikzpicture}.

Now
let boundaries $X,Y,Z,T$ and cowordisms
$\sigma:X\to Y$, $\tau:Z\to T$ be given.
Let us write $\sigma_0$, respectively $\tau_0$, for the regular part of (the underlying multiword of) $\sigma$, respectively $\tau$, and let us write $\sigma_c$, $\tau_c$ for the respective singular parts.

The {\it tensor product cowordism} $\sigma\otimes \tau:X\otimes Z\to Y\otimes T$ is  defined by the multiword
 with the   singular part
  $(\sigma\otimes \tau)_c=\sigma_c+ \tau_c$, and the regular part $(\sigma\otimes \tau)_0$
  obtained as the union of edge sets $\sigma_0$, $\tau_0$ appropriately shifted:
 $$(\sigma\otimes \tau)_0=\{[i+|Z|,w,j+|Z|]|~[i,w,j]\in \sigma_0\}\cup$$
 $$\{[\mbox{shift}_{|Z|+1}^{|X|+|Y|}(i),w,\mbox{shift}_{|Z|+1}^{|X|+|Y|}(j)]|~[i,w,j]\in \tau_0\}.$$

\begin{figure}[t]
\begin{subfigure}{.3\textwidth}
\begin{tikzpicture}[scale=.7]
    \node[left] at(-1,-0.125){$\sigma\otimes\tau:$};
    \draw[thick](0,.25)--(-0.5,.25);

    \node[left] at(-0.5,.25){$X$};

    \draw[thick](1,.25)--(1.5,.25);
    \node[right] at(1.5,0.25){$Y$};

    \draw[draw=black,fill=gray!10](0.,0)rectangle(1,.5);\node at(.5,.25){$\sigma$};

    \begin{scope}[shift={(0,-.75)}]

    \draw[thick](0,.25)--(-0.5,.25);

    \node[left] at(-0.5,.25){$Z$};

    \draw[thick](1,.25)--(1.5,.25);
    \node[right] at(1.5,0.25){$T$};

    \draw[draw=black,fill=gray!10](0.,0)rectangle(1,.5);\node at(.5,.25){$\tau$};
    \end{scope}
    \end{tikzpicture}
    \caption{Tensor product}
    \label{tensor}
    \end{subfigure}
\begin{subfigure}{.3\textwidth}
\begin{tikzpicture}[xscale=.7]
    \node[left] at(-1.25,0.5){$s_{XY}:$};
    \draw[thick](-0.5,.25)--(0.,.25)--(0.5,.75)--(1,.75);
    \draw[thick](-0.5,.75)--(0.,.75)--(0.5,.25)--(1,.25);

    \node[left] at(-0.5,.25){$X$};
    \node[left] at(-0.5,.75){$Y$};

    \node[right] at(1,0.75){$X$};
    \node[right] at(1,0.25){$Y$};
    \end{tikzpicture}
    \caption{Symmetries}
    \label{symmetries}
\end{subfigure}
\begin{subfigure}{.3\textwidth}
 \begin{tikzpicture}[xscale=.4,yscale=.7]
    \begin{scope}[shift={(-6.25,0)}]
    \node[left]at(-.7,.25){$\sigma:$};
    \node[left]at(0,.25){$X$};
    \begin{scope}[shift={(.2,0)}]
    \draw[thick](-.25,0.25)--(0.25,0.25);

    \node[right] at(2.5,.25){$Y$};

    \draw[draw=black,fill=gray!10](0.25,0)rectangle(2,.5);\node at(1.125,.25){$\sigma$};
    \draw[thick,-](2,0.25)--(2.5,0.25);
    \end{scope}

    \end{scope}

    \node[left]at(-.05,.25){$\sigma^\bot:$};

    \draw[thick](0.25,0.25)
    to[out=180,in=-90](-.25,.5)
    to[out=90,in=180](.25,1)--(1.6,1)
    ;
    \node at(2.2,1){$X^\bot$};

    \draw[draw=black,fill=gray!10](0.25,0)rectangle(2,.5);\node at(1.125,.25){$\sigma$};
    \draw[thick,-](2,0.25)
    to[out=0,in=90](2.5,0)
    to[out=-90,in=0](2,-0.5)--(.7,-0.5)
    ;

    \node at(.05,-0.5){$Y^\bot$};

    \end{tikzpicture}
    \caption{Duality}
    \label{duality}
\end{subfigure}
\\

\begin{subfigure}{.4\textwidth}
\scalebox{.8}
{
 \begin{tikzpicture}[scale=.5]

    \begin{scope}[shift={(-7,0)}]
    \node[left]at(-.7,.75){$\sigma:$};
    \draw[dashed](0.25,1.2)--(0,1.2)--(0,.3)--(0.25,.3);
    \node[left]at(0,.75){$X$};
    \begin{scope}[shift={(.8,0)}]
    \draw[thick,->](-.3,0.3)--(0.25,0.3);
    \draw[thick,<-](-.3,0.6)--(0.25,0.6);
    \draw[thick,->](-.3,.9)--(0.25,.9);
    \draw[thick,->](-.3,1.2)--(0.25,1.2);

    \draw [fill] (2.55,.9) circle [radius=0.05];
    \draw [fill] (-.3,0.3) circle [radius=0.05];
    \draw [fill] (-.3,.9) circle [radius=0.05];
    \draw [fill] (-.3,1.2) circle [radius=0.05];

    \draw[draw=black,fill=gray!10](0.25,0)rectangle(2,1.5);\node at(1.125,.75){$\sigma$};
    \draw[thick,->](2,0.3)--(2.55,0.3);
    \draw[thick,->](2,.6)--(2.55,0.6);
    \draw[thick,<-](2,.9)--(2.55,.9);
    \draw[thick,->](2,1.2)--(2.55,1.2);
    \end{scope}

    \begin{scope}[shift={(3.5,0)}]
    \draw[dashed](0.,1.2)--(0.25,1.2)--(0.25,.3)--(0.,.3);
    \node[right]at(0.25,.75){$Y$};
    \end{scope}
    \end{scope}

    \begin{scope}[shift={(-0.9,-1.5)}]
    \draw[dashed](0.25,1.2)--(0,1.2)--(0,.3)--(0.25,.3);
    \node[left]at(0,.75){$Y^\bot$};
    \end{scope}

    \begin{scope}[shift={(2.9,1.5)}]
    \draw[dashed](0,1.2)--(0.25,1.2)--(0.25,.3)--(0,.3);
    \node[right]at(0.25,.75){$X^\bot$};
    \end{scope}

    \node[left]at(-.55,.75){$\sigma^\bot:$};

    \draw[thick,<-](.25,0.3)
    to[out=180,in=-90](-.95,1.5)
    to[out=90,in=180](.25,2.7)--(2.55,2.7)
    ;
    \draw[thick,->](0.25,0.6)
    to[out=180,in=-90](-.65,1.5)
    to[out=90,in=180](.25,2.4)--(2.55,2.4)
    ;
    \draw[thick,<-](0.25,.9)
    to[out=180,in=-90](-.35,1.5)
    to[out=90,in=180](.25,2.1)--(2.55,2.1)
    ;
    \draw[thick,<-](0.25,1.2)
    to[out=180,in=-90](-.05,1.5)
    to[out=90,in=180](.25,1.8)--(2.55,1.8)
    ;

    \draw [fill] (2.55,2.7) circle [radius=0.05];
    \draw [fill] (2.55,2.1) circle [radius=0.05];
    \draw [fill] (2.55,1.8) circle [radius=0.05];

    \draw[draw=black,fill=gray!10](0.25,0)rectangle(2,1.5);\node at(1.125,.75){$\sigma$};
    \draw[thick,->](2,0.3)
    to[out=0,in=90](2.3,0)
    to[out=-90,in=0](2,-0.3)--(-0.3,-0.3)
    ;
    \draw[thick,->](2,.6)
    to[out=0,in=90](2.6,0)
    to[out=-90,in=0](2,-0.6)--(-0.3,-0.6)
    ;
    \draw[thick,<-](2,.9)
    to[out=0,in=90](2.9,0)
    to[out=-90,in=0](2,-0.9)--(-0.3,-0.9)
    ;
    \draw[thick,->](2,1.2)
    to[out=0,in=90](3.2,0)
    to[out=-90,in=0](2,-1.2)--(-0.3,-1.2)
    ;

    \draw [fill] (-0.3,-.9) circle [radius=0.05];
    \end{tikzpicture}
    }
    \caption{Duality in detail}
    \label{duality_detailed}
\end{subfigure}
\hspace{2em}
\begin{subfigure}{.4\textwidth}
\scalebox{.8}
{
\begin{tikzpicture}[scale=.8]
    \draw[thick](0,.25)--(-0.3,.25);

    \node[left] at(-0.3,.25){$X$};

    \draw[thick](0,.75)--(-0.3,.75);

    \node[left] at(-0.3,.75){$Y$};

    \draw[thick](1,.5)--(1.3,.5);
    \node[right] at(1.3,0.5){$Z$};

    \draw[draw=black,fill=gray!10](0.,0)rectangle(1,1);
    \node at (2,.5){$\cong$};
    \begin{scope}[shift={(3,-.6)}]
    \draw[thick](0,.25)--(-0.3,.25);

    \node[left] at(-0.3,.25){$X$};

    \draw[thick](0,.75)
    to[out=180,in=-90](-.5,1.25)
    to[out=90,in=180](0,1.75)
    --(1.3,1.75)
    ;

    \node[right] at(1.3,1.75){$Y^\bot$};
    \draw[thick](1,.5)--(1.3,.5);
    \node[right] at(1.3,0.5){$Z$};

    \draw[draw=black,fill=gray!10](0.,0)rectangle(1,1);
    \end{scope}


    \end{tikzpicture}
    }
    \caption{Compact structure}
    \label{compactness_picture}
\end{subfigure}
\caption{Structure of the cowordism category}
\end{figure}
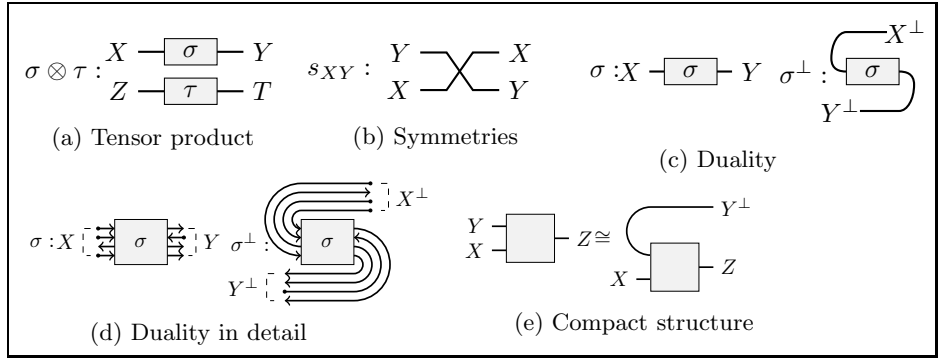

In the graphical language, tensor product of cowordisms corresponds simply to putting two boxes side by side, as in Figure \ref{tensor}.

The {\it symmetry cowordism}
$s_{X,Y}:X\otimes Y\to Y\otimes X$
is defined by the regular multiword with the
 set of edges
 $$\{[|Y|-i+1,\epsilon,|X|+|Y|+i]|~i\in Y_r\}\cup\{[|Y|+|X|-i+1,\epsilon,|X|+2|Y|+i]|~i\in X_r\}\cup$$
 $$
 \{[|X|+|Y|+i,\epsilon,|Y|-i+1]|~i\in Y_l\}\cup
\{[|X|+2|Y|+i,\epsilon,|Y|+|X|-i+1]|~i\in X_l\}.
$$
A schematic  picture of $s_{X,Y}$ is given in Figure \ref{symmetries}.


Finally, let us extend   duality from boundaries to cowordisms.

Let $X$, $Y$ be boundaries, and  $\sigma:X\to Y$ be a cowordism. Let us identify $\sigma$ with the underlying multiword $\sigma=(\sigma_0,\sigma_c)$.

The {\it dual cowordism} $\sigma^\bot:Y^\bot\to X^\bot$ of $\sigma$ is
  the multiword with the same singular part
 $\sigma_c$ and the regular part $\sigma_0^\bot$  obtained from $\sigma_0$  by
  a cyclic permutation of boundary vertices:
   $\sigma_0^\bot=\{[\phi(i),w,\phi(j)]|~[i,w,j]\in \sigma_0\}$, where
$\phi(i)=i+|Y|,\mbox{ if }i\leq |X|,~\phi(i)=i-|X|,\mbox{ if }i> |X|$.

In a schematic  picture, duality is shown in  Figure \ref{duality}.

The full picture, again, can be recovered by expanding every wire  into a parallel cluster.
For example, if $X,Y$ are as in (\ref{graph.lang.example}), the above picture translates to the one in Figure \ref{duality_detailed}.
(We defined duality to flip tensor factors precisely in order to have  this consistency with ``parallel wires substitution'' in the graphical language.)

It is very easy to check that, for a fixed alphabet $T$, we have a well-defined category ${\bf Coword_T}$ of boundaries and cowordisms, and the operation of tensor product together with symmetry cowordisms make it a {\it symmetric monoidal} category. Moreover there are natural isomorphisms
\begin{equation}\label{compact structure}
(X\otimes Y)^\bot\cong X^\bot\otimes Y^\bot\quad \mbox{Hom}(Y\otimes X,Z)\cong \mbox{Hom}(X,Y^\bot\otimes Z),
\end{equation}
which means that the duality makes the category {\it compact} (see \cite{KellyLaplaza}, also \cite{AbramskyCoecke}).
The first isomorphism in (\ref{compact structure}) is the symmetry;  the second one is shown in Figure \ref{compactness_picture}.

In fact, in a sense that can be made precise, the category of cowordisms over an alphabet $T$ is a {\it free compact} category generated by the free monoid $T^*$, where the latter is seen as a  category with one object (compare with \cite{AbramskyFreeTraced}).

\section{Representing linear $\lambda$-calculus}
Here we assume that the reader is familiar with basic notion of $\lambda$-calculus, see \cite{Barendregt} for reference.

We assume that we are given   sets $X$ and $C$ of {\it variables} and   {\it constants}, with $C\cap X=\emptyset$. The set $\Lambda=\Lambda(X,C)$ of {$\lambda$-terms} is constructed from $X$ and $C$ by applications and $\lambda$-abstractions.

In linear $\lambda$-calculus, terms are typed using (intuitionistic) {\it implicational linear logic} ({\bf ILL}).

Given a  set $N$ of {\it literals} or {\it atomic types}, the set  $Tp=Tp(N)$  of
 {\it linear implicational types} (over $N$),   is defined by the grammar
$Tp::=N|Tp\multimap Tp$.

A {\it typing judgement} is a sequent of the form
$x_1:A_1,\ldots,x_n:A_n\vdash t:A$,
where $x_1,\ldots x_n\in X$ are pairwise distinct ($n$ may be zero), $t\in\Lambda(X,C)$, and $A_1,\ldots,A_n,A\in Tp(N)$.

  \begin{figure}
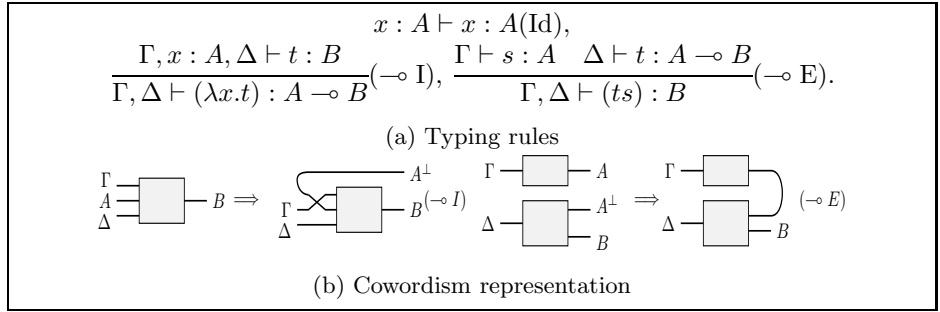

 \begin{subfigure}{1\textwidth}
 \centering
 ${x:A\vdash x:A}(\mbox{Id})$,\\
$\cfrac{\Gamma,x:A,\Delta\vdash t:B}{\Gamma,\Delta\vdash (\lambda x.t):A\multimap B}(\multimap\mbox{I})$,
$\cfrac{\Gamma\vdash s:A\quad\Delta\vdash t:A\multimap B}{\Gamma,\Delta\vdash (ts):B}(\multimap\mbox{E})$.
\caption{Typing rules}
\label{lambda-rules}
 \end{subfigure}
 \\
\begin{subfigure}{1\textwidth}
 \centering
 \scalebox{.6}[.8]
 {
  \tikz[scale=.5]
         {
             \begin{scope}[shift={(1,0)}]
             \begin{scope}[shift={(.5,0)}]
                     \draw[draw=black,fill=gray!10](0,.55)rectangle(2,2.05);
             \draw[thick,-](-1,1.8)--(0,1.8);
             \draw[thick,-](-1,1.3)--(0,1.3);
             \draw[thick,-](-1,.8)--(0,.8);

             \node[left] at(-1,2) {$\Gamma$};
             \node[left] at(-1,1.3) {$A$};
             \node[left] at(-1,.6) {$\Delta$};

            \draw[thick,-](2,1.3)--(3,1.3);

             \node[right] at(3,1.3) {$B$};
             \end{scope}
             \node at(5.25,1.3) {$\Longrightarrow$};
             \end{scope}

            \begin{scope}[shift={(2.25,0)}]
                     \draw[draw=black,fill=gray!10](8,.25)rectangle(10,1.75);
             \draw[thick,-](6.25,1)--(6.75,1.)--(7.5,1.5)--(8,1.5);
             \draw[thick,-](6.75,1.5)--(7.5,1.)--(8,1.);
             \draw[thick,-](6.25,.5)--(8,.5);
            \draw[thick,-](6.75,1.5)
            to[out=180,in=-90](6.25,2.)
            to[out=90,in=180](6.75,2.25)
            --(11,2.25);

             \node[left] at(6.25,1) {$\Gamma$};
             \node[right] at(11,2.25) {$A^\bot$};
             \node[left] at(6.25,.3) {$\Delta$};

            \draw[thick,-](10,1)--(11,1);

             \node[right] at(11,1) {$B$};
             \node[left] at(14,1.3) {$(\multimap I)$};
            \end{scope}

            \begin{scope}[shift={(18.5,1.8)}]
            \draw[draw=black,fill=gray!10](0,0)rectangle(2,1);
             \draw[thick,-](2,.5)--(3,.5);

             \node[right] at(3,.5) {$A$};
             \draw[thick,-](-1,.5)--(0,.5);

             \node[left] at(-1,.5) {$\Gamma$};

                \draw[draw=black,fill=gray!10](0,-2)rectangle(2,-.5);
            \draw[thick,-](2,-.8)--(3,-.8);

             \node[right] at(3,-.6) {$A^\bot$};
            \draw[thick,-](2,-1.7)--(3,-1.7);
              \node[right] at(3,-1.9) {$B$};

             \draw[thick,-](-1,-1.25)--(0,-1.25);
             \node[left] at(-1,-1.25) {$\Delta$};

            \node at(5.5,-.5) {$\Longrightarrow$};
            \begin{scope}[shift={(1,0)}]
                     \draw[draw=black,fill=gray!10](7,0)rectangle(9,1);
             \draw[thick,-](9,.5)--(10,.5)
             to[out=0,in=90](10.5,-.25)
             to[out=-90,in=0](10,-1)
             ;

             \draw[thick,-](6,.5)--(7,.5);

             \node[left] at(6,.5) {$\Gamma$};

                     \draw[draw=black,fill=gray!10](7,-2)rectangle(9,-.5);
            \draw[thick,-](9,-1)--(10 ,-1);

            \draw[thick,-](9,-1.5)--(10,-1.5);
              \node[right] at(10,-1.5) {$B$};

             \draw[thick,-](6,-1.25)--(7,-1.25);
             \node[left] at(6,-1.25) {$\Delta$};

            \node[right] at(11,-.5) {$(\multimap E)$};

            \end{scope}

            \end{scope}

         }
         }
            \caption{Cowordism representation}
\end{subfigure}
\caption{Cowordism representation of linear $\lambda$-calculus }
\label{lambda-interpretation}
\end{figure}

A {\it linear signature}, or, simply, a {\it signature}, $\Sigma$ is a triple $\Sigma=(N,C,\mathfrak{T})$, where $N$ is a finite set of atomic types, $C$ is a finite set of constants and $\mathfrak{T}$ is a function assigning to each constant $c\in C$ a linear implicational type $\mathfrak{T}(c)\in Tp(N)$. We say that $\Sigma$ is a signature {\it over the set $N$ of atomic types}.

Typing judgements of the form
$\vdash c:\mathfrak{T}(c)$,
 where $c\in C$, are called {\it signature axioms} of $\Sigma$.

 Typing judgements are derived using type inference rules in Figure \ref{lambda-rules} (which happen to be {\it natural deduction} rules of {\bf ILL} decorated with $\lambda$-terms).
Given a signature  $\Sigma$, we say that a typing judgement is {\it derivable in} $\Sigma$ if it is derivable from axioms of $\Sigma$ by rules of linear $\lambda$-calculus. We write in this case $\Gamma\vdash_\Sigma t:A$.

It is well known \cite{HylandDePaiva}
that any {\it symmetric monoidal closed category}, in particular, a compact closed category, provides a denotational model for linear $\lambda$-calculus (invariant under $\beta\eta$-equivalence). We specialize to the concrete case of the category ${\bf Coword_T}$ of cowordisms over the given alphabet $T$.

So, let the sets $N$  and $T$ of literals and terminal symbols respectively be given.  An {\it interpretation $\xi$ of linear types} over $N$  in ${\bf Coword_T}$ consists in assigning to each atomic type $p\in N$ a boundary $\xi(p)$. This is extended to all types in $Tp(N)$ by
$\xi(A\multimap B)=\xi(A)^\bot\otimes\xi(B)$.

Now, given a linear signature $\Sigma$ over $N$ and $T$,  we want to extend the interpretation  to derivable typing judgements, so that a judgement of the form
$x_1:A_1,\ldots,x_n:A_n\vdash t:A$ is interpreted as a cowordism of the form
            $\xi(A_1)\otimes\ldots\otimes \xi(A_n)\to \xi(A)$.

 {\it Interpretation of typing judgements} consists in assigning, for each constant $c$ and axiom $\vdash c:A$ of $\Sigma$ (here $A=\mathfrak{T}(c)$), a multiword $\xi(c)$ with boundary $\xi(A)$, which we identify with a cowordism  $\xi(c):{\bf 1}\to \xi(A)$.

 This is extended to all typing judgements derivable in $\Sigma$ by induction on type inference rules.
The (Id) axiom
 $x:A\vdash x:A$  is  interpreted as the identity cowordism $\id_{\xi(A)}$. Typing judgements obtained by
 the ($\multimap\mbox{I}$) or ($\multimap\mbox{E}$) rules are interpreted according to Figure \ref{lambda-interpretation} (where the symbol $\xi$ is omitted).

In the sequel we often will  abuse notation and denote a type in $Tp(N)$ and its interpretation in ${\bf Coword_T}$ with the same symbol, as is customary in the literature.

\section{String abstract categorial grammars}
The {\it string signature} $Str_T$ {\it over} $T$, where  $T$ is a finite alphabet, is the linear signature with a single atomic type $O$, the alphabet $T$ as the set of constants and the  typing assignment
$\mathfrak{T}(c)=O\multimap O~\forall c\in T.$
We denote the type $O\multimap O$ as $str$.

Any word $w=a_1\ldots a_n$ in the alphabet $T$ can be represented as the  term
$\rho(w)=a_1\circ\ldots\circ a_n$,
where
$a_1\circ\ldots\circ a_n=(\lambda x.a_1(\ldots(a_n(x))\ldots))$,
and $\vdash_{Str_T} \rho(w):str$.

Moreover, it can be shown that, for any term $t$, if $\vdash_{Str_T} t:str$ then $t\sim_{\beta\eta}\rho(w)$ for some $w\in T^*$.

{\it The cowordism representation} $\xi_0$ of the string signature $Str_T$ over the alphabet $T$ is given by the following interpretation in ${\bf Coword_T}$.

   For the  atomic type $O$ we put
   $\xi_0(O)=\{1\}$, $(\xi_0(O))_l=\emptyset$. (I.e. $\xi_0(O)$ is a single-point boundary).

    Then for each axiom $\vdash c:O\multimap O$, where $c\in T$, we put $\xi_0(c)=[1,c,2]$. The latter is the multiword with boundary $O^\bot\otimes O$  consisting of a single edge labeled with $c$:
    \begin{tikzpicture}[baseline=-8pt,xscale=1.5,yscale=0.7]
            \draw [fill] (0,0) circle [radius=0.05];
            \draw[thick,->] (0,0) to  [out=180,in=90] (-0.5,-0.25) to  [out=-90,in=180] (0,-.5);
            \node[ left] at (-0.5,-0.3) {$c$};
        \end{tikzpicture}.

Given  linear signatures $\Sigma_i=(N_i,C_i,\mathfrak{T}_i)$, $i=1,2$, a {\it homomorphism of signatures} $\phi:\Sigma_1\to\Sigma_2$ is a
pair
of maps
$$\phi_{Tp}:Tp(N_1)\to  Tp(N_2),\quad   \phi_{Tm}:\Lambda(X,C_1)\to\Lambda(X,C_2),$$  such that
$\phi_{Tp}(A\multimap
      B)=\phi_{Tp}(A)\multimap \phi_{Tp}(B)$,
       $\phi_{Tm}(ts)=(\phi_{Tm}(t)\phi_{Tm}(s))$,
       $\phi_{Tm}(\lambda x.t)=(\lambda x.\phi_{Tm}(t))$, $\phi_{Tm}(x)=x${ for }$x${ a variable},
and      for any $c\in C_1$ it holds that $\vdash_{\Sigma_2}\phi_{Tm}(c):\phi_{Tp}(\mathfrak{T}(c))$.

An {\it  abstract categorial grammar over string signature (string ACG)} $G$ is a tuple $G=(\Sigma_{abstr},T, \phi,S)$, where
 $\Sigma_{abstr}$, the {\it abstract} signature, is a linear signature, $T$ is a finite alphabet of {\it terminal symbols},
     $\phi:\Sigma_{abstr}\to Str_T$, the {\it lexicon}, is a homomorphism of signatures, and
   $S$, {\it the initial type}, is an atomic type of $\Sigma_{abstr}$ with $\phi_{Tp}(S)=str$.
We say that  $G$ is a {string ACG} {\it over}  $T$.

The {\it string language} $L(G)$ generated by a string ACG $G$ is the set of words
$L(G)=\{w\in T^*|~\exists t~\phi_{Tm}(t)\sim_{\beta\eta}\rho(w)\&\vdash_{\Sigma_{abstr}}t:S\}$.

In the setting as above, the cowordism representation $\xi_0$ of  $Str_T$ immediately gives us  an interpretation $\xi$ of the abstract signature $\Sigma$ in the category ${\bf Coword_T}$ of cowordisms over $T$, obtained as the composition $\xi=\xi_0\circ\phi$.

That is, for any type $A\in Tp(\Sigma)$ we put
$\xi(A)=\xi_0(\phi_{Tp}(A))$,
and for any signature axiom $\vdash c:\mathfrak{T}(c)$ of $\Sigma$
we put
$\xi(c)=\xi_0(\phi_{Tm}(c))$. The latter is a multiword with boundary  $\xi(\phi_{Tp}(\mathfrak{T}(c)))=\xi_0(\mathfrak{T}(c))$.

Because $\phi$ is a homomorphism of signatures, an easy induction on derivation shows that for any
   typing judgement
$
x_1:A_1,\ldots,x_n:A_n\vdash t:A
$
derivable in $\Sigma$,
  its interpretation coincides with the interpretation of
the typing judgement
$
x_1:\phi_{Tp}(A_1),\ldots,x_n:\phi_{Tp}(A_n)\vdash \phi_{Tm}(t):\phi_{Tp}(A)
$
(which is derivable in $Str_T$).

  In particular, for the initial type $S$ we have $\xi(S)=\xi_0(O^\bot\otimes O)$ is a two-point boundary, and any derivable typing judgement of the form $\vdash_\Sigma t:S$ is interpreted as a single-edge  multiword,  labeled with  $\rho(\phi_{Tm}(t))$ .

    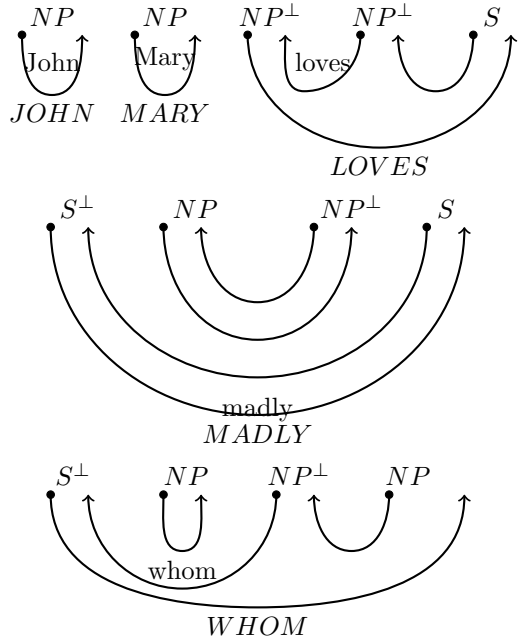
\begin{figure}[p]
    \centering
    \begin{subfigure}[h]{1\textwidth}
     \centering
        {
            \text
                {
                $\vdash JOHN:NP,\quad \vdash MARY:NP,\quad
                \vdash LOVES: NP\multimap NP\multimap S$,
                }
        }

        {
$\vdash MADLY:(NP\multimap S)\multimap NP\multimap S,$
\\
$
\vdash WHOM:(NP\multimap S)\multimap  NP\multimap NP$.
}
\caption
            {
            Axioms
            }
\label{ACG_ex_axioms}
\end{subfigure}
\\
\begin{subfigure}[h]{1\textwidth}
 \centering
        {
$\phi_{Tp}(NP)=\phi_{Tp}(S)=O\multimap O$,
}\\
        {
$\phi_{Tm}(JOHN)=\mbox{John},\quad\phi_{Tm}(MARY)=\mbox{Mary},\quad\phi_{Tm}(JIM)=\mbox{Jim}$,
}\\
        {
$\phi_{Tm}(LOVES)=\lambda xy.(y\circ \mbox{loves}\circ x),\quad
\phi_{Tm}(MADLY)=\lambda fx.((f\cdot x)\circ (\mbox{madly}))$,
}\\
        {
$
\phi_{Tm}(WHOM
)=\lambda fx.(x\circ (\mbox{whom})\circ (f\cdot(\lambda y.y))).$
}
\caption
            {
            Lexicon
            }
\label{ACG_ex_lexicon}
\end{subfigure}
\begin{subfigure}{1\textwidth}
 \centering
\begin{tikzpicture}
         \draw[thick,->](0,0) to  [out=-90,in=180] (0.4,-0.8) to  [out=0,in=-90] (.8,0);
         \draw [fill] (0,0) circle [radius=0.05];
          \node[above ] at (0.4,0) {$NP$};
           \node[above] at (0.4,-.6) {$\mbox{John}$};
        \node[below] at (0.4,-0.8){$JOHN$};

        \begin{scope}[shift ={(1.5,0)}]
         \draw[thick,->](0,0) to  [out=-90,in=180] (0.4,-0.8) to  [out=0,in=-90] (0.8,0);
         \draw [fill] (0,0) circle [radius=0.05];
          \node[above ] at (0.4,0) {$NP$};
           \node[above] at (0.4,-.6) {$\mbox{Mary}$};
        \node[below]  at (0.4,-0.8){$MARY$};
        \end{scope}

        \begin{scope}[shift ={(3,0)}]
        \draw[thick,->](0,0) to  [out=-90,in=180] (1.75,-1.5) to  [out=0,in=-90] (3.5,0);
         \draw [fill] (0,0) circle [radius=0.05];
        \draw[thick,<-](0.5,0) to  [out=-90,in=180] (0.75,-0.75) to  [out=0,in=-90] (1.5,0);
        \draw [fill] (3,0) circle [radius=0.05];
        \draw[thick,<-](2,0) to  [out=-90,in=180] (2.5,-0.75) to  [out=0,in=-90] (3,0);
        \draw [fill] (1.5,0) circle [radius=0.05];
        \node[above] at (1,-.6) {$\mbox{loves}$};
          \node[above] at (0.3,0) {$NP^\bot$};
        \node[above] at (1.8,0) {$NP^\bot$};
        \node[above ] at (3.25,0) {$S$};
        \node[below]  at(1.75,-1.5){$LOVES$};

        \end{scope}
        \end{tikzpicture}
        \\
        \begin{tikzpicture}
             \begin{scope}[yscale=1]
             \draw[thick,->](0,0) to  [out=-90,in=180] (2.75,-2.5) to  [out=0,in=-90] (5.5,0);
            \draw[thick,<-](0.5,0) to  [out=-90,in=180] (2.75,-2) to  [out=0,in=-90] (5,0);
            \draw [fill] (0,0) circle [radius=0.05];
            \draw [fill] (5,0) circle [radius=0.05];

            \draw[thick,->](1.5,0) to  [out=-90,in=180] (2.75,-1.5) to  [out=0,in=-90] (4,0);
            \draw[thick,<-](2,0) to  [out=-90,in=180] (2.75,-1) to  [out=0,in=-90] (3.5,0);
            \node[above] at (2.75,-2.7) {$\mbox{madly}$};
            \draw [fill] (1.5,0) circle [radius=0.05];
            \draw [fill] (3.5,0) circle [radius=0.05];

              \node[above right] at (0,0) {$S^\bot$};
            \node[above left] at (5.5,0) {$S$};
            \node[above right] at (1.5,0) {$NP$};
            \node[above] at (4,0) {$NP^\bot$};
            \node at (2.75,-2.75){$MADLY$};
        \end{scope}

        \end{tikzpicture}
        \\
        \begin{tikzpicture}
             \begin{scope}[yscale=1]
    \draw[thick,->](0,0) to  [out=-90,in=180] (2.75,-1.5) to  [out=0,in=-90] (5.5,0);
    \draw [fill] (0,0) circle [radius=0.05];
    \node[above] at (0.3,0) {$S^\bot$};

     \draw[thick,->](1.5,0) to  [out=-90,in=180] (1.75,-.75) to  [out=0,in=-90] (2,0);
     \draw [fill] (1.5,0) circle [radius=0.05];
    \draw[thick,<-](0.5,0) to  [out=-90,in=180] (1.75,-1.25) to  [out=0,in=-90] (3,0);
    \draw [fill] (3,0) circle [radius=0.05];
    \draw[thick,<-](3.5,0) to  [out=-90,in=180] (4,-0.75) to  [out=0,in=-90] (4.5,0);
    \draw [fill] (4.5,0) circle [radius=0.05];
    \node[above] at (1.75,-1.25) {$\mbox{whom}$};
      \node[above] at (1.8,0) {$NP$};
    \node[above] at (3.3,0) {$NP^\bot$};
    \node[above ] at (4.75,0) {$NP$};
    \node[below] at  (2.75,-1.5){$WHOM$};
        \end{scope}

        \end{tikzpicture}
        \caption{Cowordism representation}
\label{ACG_ex_cowordisms}
\end{subfigure}
\caption{Cowordism representation  of a string ACG}
\end{figure}
We give a concrete example of a string ACG and its cowordism representation.

We consider the set of atomic types $\{NP,S\}$ and the terminal alphabet
$\{\mbox{John},\mbox{Mary},\mbox{loves},\mbox{madly},\mbox{whom}\}$.

The signature axioms and the lexicon are collected in Figures \ref{ACG_ex_axioms}, \ref{ACG_ex_lexicon}, while the translation to cowordisms is shown in Figure \ref{ACG_ex_cowordisms}. We rotated pictures of cowordisms $90^{\circ}$ counterclockwise, so that
   outgoing boundaries are shown on the top, with the ordering of vertices   from left to right.

   We generate the noun phrase
$\mbox{``Mary whom John loves madly''}$,
 represented as a term of type $NP$.
The derivation is shown in Figure \ref{ACG_ex_derivation};
for convenience, we break  it into five consecutive  steps.
A step-by-step translation into the language of cowordisms is shown in Figures \ref{ACG_ex_picture}, \ref{ACG_ex_picture_cont} with omission of  the last step, which should become clear by the end.

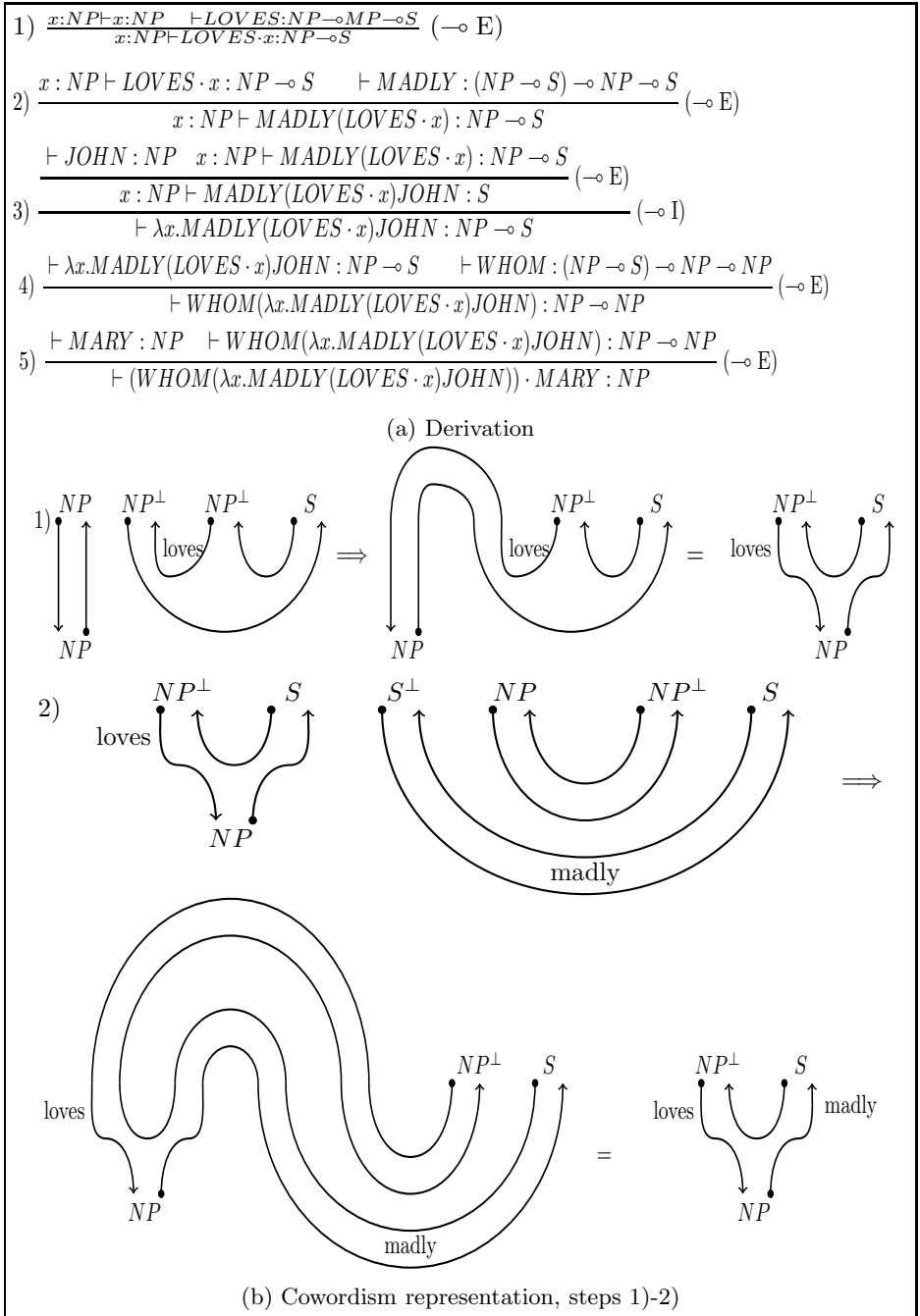
\begin{figure}
\begin{subfigure}{1\textwidth}
         {
         1)
        $\frac{x:NP\vdash x:NP\quad \vdash LOVES:NP\multimap MP\multimap S}{x:NP\vdash LOVES\cdot x:NP\multimap S}~(\multimap\mbox{E})$
        }\\
\mbox{} \\
     \scalebox{.75}[1]
         {
         2)
         $\cfrac
                {
                  x:NP\vdash LOVES\cdot x:NP\multimap S
                \quad\quad
                                                            \vdash MADLY: (NP\multimap S)\multimap NP\multimap S
                }
                {
                 x:NP\vdash MADLY(LOVES\cdot x):NP\multimap S
                }~(\multimap\mbox{E})$
        }\\
\mbox{}\\
    \scalebox{.75}[1]
        {
        3)
            $\cfrac
                {
                    \cfrac
                    {
                    \vdash JOHN:NP\quad
                     x:NP\vdash MADLY(LOVES\cdot x):NP\multimap S
                    }
                    {
                    x:NP\vdash MADLY(LOVES\cdot x)JOHN: S
                    }~(\multimap\mbox{E})
                }
                {
                    \vdash \lambda x.MADLY(LOVES\cdot x)JOHN: NP\multimap S
                }~(\multimap\mbox{I})$
        }\\
\mbox{} \\
      \scalebox{.7}[1]
         {
         4)
                $\cfrac{
                \vdash \lambda x.MADLY(LOVES\cdot x)JOHN:NP\multimap S\quad\quad
                \vdash WHOM:(NP\multimap S)\multimap NP\multimap NP
                }
                {
                \vdash WHOM(\lambda x.MADLY(LOVES\cdot x)JOHN):NP\multimap NP
                }~(\multimap\mbox{E})$
         }\\
\mbox{} \\
      \scalebox{.75}[1]
         {
         5)
                $\cfrac
                {
                \vdash MARY:NP\quad \vdash WHOM(\lambda x.MADLY(LOVES\cdot x)JOHN):NP\multimap NP
                }
                {
                \vdash (WHOM(\lambda x.MADLY(LOVES\cdot x)JOHN))\cdot MARY:NP
                }~(\multimap\mbox{E})$
        }\\
\caption{Derivation}
\label{ACG_ex_derivation}
\end{subfigure}
\begin{subfigure}{1\textwidth}
\centering
\scalebox{.75}[1.]
{
 \begin{tikzpicture}[xscale=1]
\begin{scope}[shift={(-1.5,0)}]
 \node [left] at(0,0){1)};
\draw[thick,->](0,0) -- (0,-1.5);
\draw[thick,<-](0.5,0) -- (0.5,-1.5);
 \node[above] at (0.3,0) {$NP$};
  \node[below] at (0.3,-1.5) {$NP$};
  \draw [fill] (0,0) circle [radius=0.05];
  \draw [fill] (0.5,-1.5) circle [radius=0.05];
\end{scope}

\begin{scope}[shift={(-.25,0)}]
     \draw[thick,->](0,0) to  [out=-90,in=180] (1.75,-1.5) to  [out=0,in=-90] (3.5,0);

     \draw [fill] (0,0) circle [radius=0.05];
    \draw[thick,<-](0.5,0) to  [out=-90,in=180] (0.75,-0.75) to  [out=0,in=-90] (1.5,0);
    \draw [fill] (1.5,0) circle [radius=0.05];



    \node[above] at (1,-.6) {$\mbox{loves}$};
      \node[above] at (0.3,0) {$NP^\bot$};
      \node[above] at (1.8,0) {$NP^\bot$};
      \node[above] at (3.3,0) {$S$};

    \draw[thick,<-](2,0) to  [out=-90,in=180] (2.5,-0.75) to  [out=0,in=-90] (3,0);
    \draw [fill] (3,0) circle [radius=0.05];
\end{scope}

{
\node at(3.8,-.5){$\Longrightarrow$};
}

\begin{scope}[shift={(6.,0)}]

    \begin{scope}[shift={(-1.5,0)}]
    \draw[thick,->](0,0) -- (0,-1.5);
    \draw[thick,-](0.5,0) -- (0.5,-1.5);
      \node[below ] at (0.3,-1.5) {$NP$};
      \draw [fill] (0.5,-1.5) circle [radius=0.05];
    \end{scope}

    \draw[thick](0,0) to  [out=90,in=0] (-.75,.5) to  [out=180,in=90] (-1.,0);
    \draw[thick](0.5,0) to  [out=90,in=0] (-.75,1) to  [out=180,in=90] (-1.5,0);

     \draw[thick,->](0,0) to  [out=-90,in=180] (1.75,-1.5) to  [out=0,in=-90] (3.5,0);

    \draw[thick,-](0.5,0) to  [out=-90,in=180] (0.75,-0.75) to  [out=0,in=-90] (1.5,0);
    \draw [fill] (1.5,0) circle [radius=0.05];



    \node[above] at (1,-.6) {$\mbox{loves}$};
      \node[above] at (1.8,0) {$NP^\bot$};
      \node[above] at (3.3,0) {$S$};

    \draw[thick,<-](2,0) to  [out=-90,in=180] (2.5,-0.75) to  [out=0,in=-90] (3,0);
    \draw [fill] (3,0) circle [radius=0.05];

\end{scope}
\node [font=\large] at(10.,-.5) {
$=$
};
\begin{scope}[shift={(10,0)}]

    \begin{scope}
     \draw[thick,->](2.75,-1.5) to  [out=90,in=180] (3.25,-.75) to  [out=0,in=-90] (3.5,0);

     \draw [fill] (2.75,-1.5) circle [radius=0.05];
    \draw[thick,<-](2.25,-1.5) to  [out=90,in=0] (1.75,-0.75) to  [out=180,in=-90] (1.5,0);
    \draw [fill] (1.5,0) circle [radius=0.05];



    \node[above] at (1,-.6) {$\mbox{loves}$};
      \node[below] at (2.45,-1.5) {$NP$};
      \node[above] at (1.8,0) {$NP^\bot$};
      \node[above] at (3.3,0) {$S$};

    \draw[thick,<-](2,0) to  [out=-90,in=180] (2.5,-0.75) to  [out=0,in=-90] (3,0);
    \draw [fill] (3,0) circle [radius=0.05];
    \end{scope}

\end{scope}
\end{tikzpicture}
}\\
%
{
    \begin{tikzpicture}[xscale=1]
    \node at(0,0){2)};
     \draw[thick,->](2.75,-1.5) to  [out=90,in=180] (3.25,-.75) to  [out=0,in=-90] (3.5,0);

     \draw [fill] (2.75,-1.5) circle [radius=0.05];
     \draw[thick,->] (4.5,0) to  [out=-90,in=180] (7.25,-2.5) to  [out=0,in=-90] (10,0);
     \draw [fill] (4.5,0) circle [radius=0.05];

    \draw[thick,<-](2.25,-1.5) to  [out=90,in=0] (1.75,-0.75) to  [out=180,in=-90] (1.5,0);
    \draw [fill] (1.5,0) circle [radius=0.05];
     \draw[thick,<-](6.5,0) to  [out=-90,in=180] (7.25,-1) to  [out=0,in=-90] (8,0);


    \node[above] at (1,-.6) {$\mbox{loves}$};
        \node[below] at (2.45,-1.5) {$NP$};
      \node[above] at (1.8,0) {$NP^\bot$};
      \node[above] at (3.3,0) {$S$};
    \node[above] at (4.8,0) {$S^\bot$};
    \node[above] at (6.3,0) {$NP$};

    \draw[thick,<-](2,0) to  [out=-90,in=180] (2.5,-0.75) to  [out=0,in=-90] (3,0);
    \draw [fill] (3,0) circle [radius=0.05];
     \draw[thick,<-](5,0) to  [out=-90,in=180] (7.25,-2) to  [out=0,in=-90] (9.5,0);

    \draw [fill] (9.5,0) circle [radius=0.05];

    \draw [fill] (6,0) circle [radius=0.05];
    \draw[thick,->](6,0) to  [out=-90,in=180] (7.25,-1.5) to  [out=0,in=-90] (8.5,0);
    \node[above] at (7.25,-2.5) {$\mbox{madly}$};
    \draw [fill] (8,0) circle [radius=0.05];

    \node[above left] at (10,0) {$S$};

    \node[above] at (8.5,0) {$NP^\bot$};

\node at (11,-1){$\Longrightarrow$};
    \end{tikzpicture}
    }
\\

\scalebox{.75}[1.]
{
    \begin{tikzpicture}[xscale=1]
     \draw[thick,-](2.75,-1.5) to  [out=90,in=180] (3.25,-.75) to  [out=0,in=-90] (3.5,0);

     \draw [fill] (2.75,-1.5) circle [radius=0.05];
     \draw[thick,->] (4.5,0) to  [out=-90,in=180] (7.25,-2.5) to  [out=0,in=-90] (10,0);

    \draw[thick,<-](2.25,-1.5) to  [out=90,in=0] (1.75,-0.75) to  [out=180,in=-90] (1.5,0);
     \draw[thick,-](6.5,0) to  [out=-90,in=180] (7.25,-1) to  [out=0,in=-90] (8,0);


    \node[above] at (1,-.6) {$\mbox{loves}$};
        \node[below] at (2.45,-1.5) {$NP$};

    \draw[thick,-](2,0) to  [out=-90,in=180] (2.5,-0.75) to  [out=0,in=-90] (3,0);
     \draw[thick,-](5,0) to  [out=-90,in=180] (7.25,-2) to  [out=0,in=-90] (9.5,0);

    \draw [fill] (9.5,0) circle [radius=0.05];

    \draw[thick,->](6,0) to  [out=-90,in=180] (7.25,-1.5) to  [out=0,in=-90] (8.5,0);
    \node[above] at (7.25,-2.5) {$\mbox{madly}$};
    \draw [fill] (8,0) circle [radius=0.05];

    \node[above left] at (10,0) {$S$};

    \node[above] at (8.5,0) {$NP^\bot$};

    \draw[thick,-](3.5,0) to [out=90,in=180] (4,0.5) to [out=0,in=90](4.5,0);
    \draw[thick,-](1.5,0)to [out=90,in=180] (4,2.5) to [out=0,in=90](6.5,0);
    \draw[thick,-](3,0) to [out=90,in=180] (4,1) to [out=0,in=90](5.,0);
    \draw[thick,-](2,0)to [out=90,in=180] (4,2.) to [out=0,in=90](6.,0);

\node at (10.75,-1){$=$};
\begin{scope}[shift={(11,0)}]
    \begin{scope}
     \draw[thick,->](2.75,-1.5) to  [out=90,in=180] (3.25,-.75) to  [out=0,in=-90] (3.5,0);

     \draw [fill] (2.75,-1.5) circle [radius=0.05];
    \draw[thick,<-](2.25,-1.5) to  [out=90,in=0] (1.75,-0.75) to  [out=180,in=-90] (1.5,0);
    \draw [fill] (1.5,0) circle [radius=0.05];



    \node[above] at (1,-.6) {$\mbox{loves}$};
    \node[above] at (4.2,-.6) {$\mbox{madly}$};
      \node[below] at (2.45,-1.5) {$NP$};
      \node[above] at (1.8,0) {$NP^\bot$};
      \node[above] at (3.3,0) {$S$};

    \draw[thick,<-](2,0) to  [out=-90,in=180] (2.5,-0.75) to  [out=0,in=-90] (3,0);
    \draw [fill] (3,0) circle [radius=0.05];
    \end{scope}

\end{scope}
\end{tikzpicture}
}
\caption{Cowordism representation, steps 1)-2)}
\label{ACG_ex_picture}
\end{subfigure}
\caption{ACG representation example}
\end{figure}
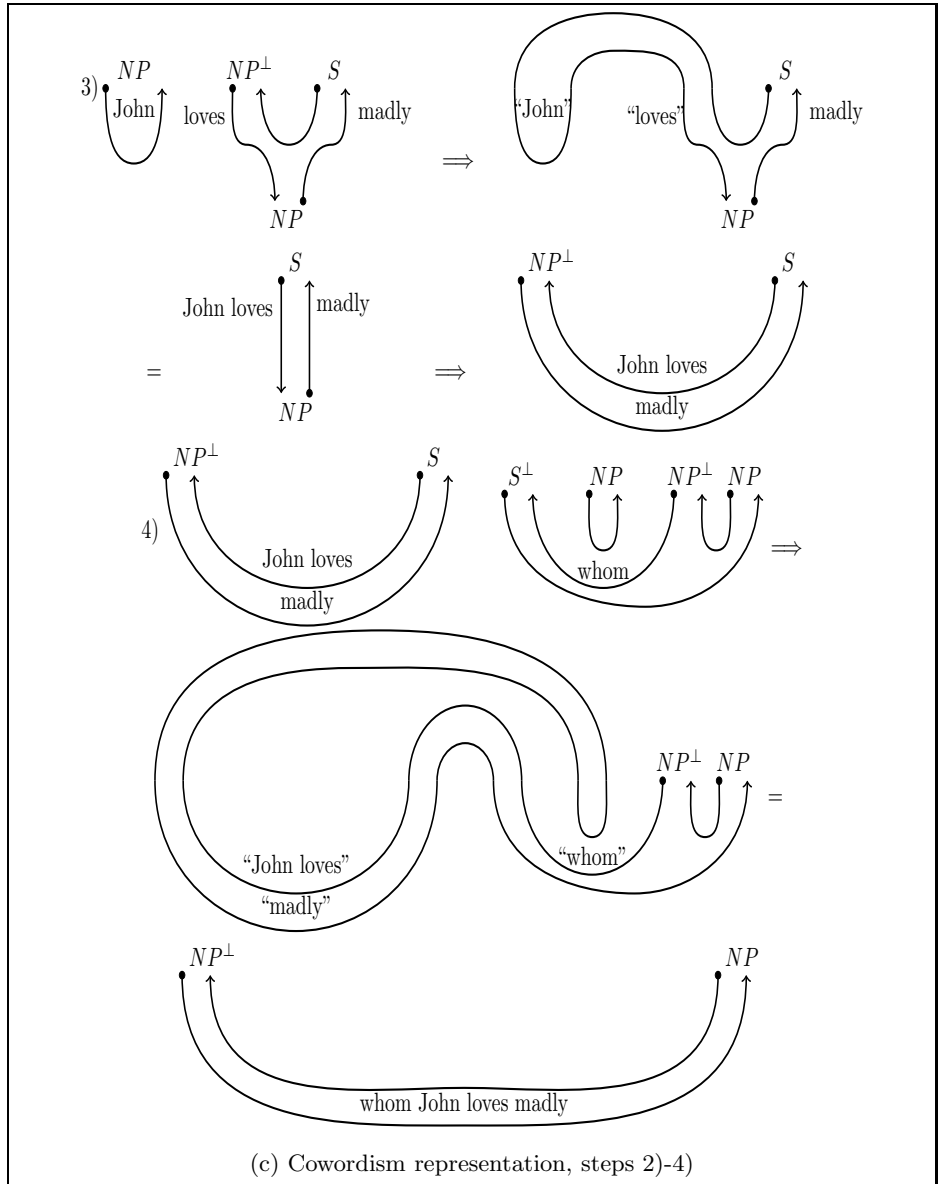
\begin{figure}\ContinuedFloat
\begin{subfigure}[b]{1\textwidth}
\centering
\scalebox{.75}[1.]
{
    \begin{tikzpicture}

        \begin{scope}

         \draw[thick,->](2.75,-1.5) to  [out=90,in=180] (3.25,-.75) to  [out=0,in=-90] (3.5,0);

         \draw [fill] (2.75,-1.5) circle [radius=0.05];
        \draw[thick,<-](2.25,-1.5) to  [out=90,in=0] (1.75,-0.75) to  [out=180,in=-90] (1.5,0);
        \draw [fill] (1.5,0) circle [radius=0.05];



        \node[above] at (1,-.6) {$\mbox{loves}$};
        \node[above] at (4.2,-.6) {$\mbox{madly}$};
          \node[below] at (2.45,-1.5) {$NP$};
          \node[above] at (1.8,0) {$NP^\bot$};
          \node[above] at (3.3,0) {$S$};

        \draw[thick,<-](2,0) to  [out=-90,in=180] (2.5,-0.75) to  [out=0,in=-90] (3,0);
        \draw [fill] (3,0) circle [radius=0.05];
        \end{scope}

        \begin{scope}[shift={(-.75,0)}]
        \node [left] at (0,0){3)};
         \draw[thick,->](0,0) to  [out=-90,in=180] (0.5,-1) to  [out=0,in=-90] (1,0);
         \draw [fill] (0,0) circle [radius=0.05];
          \node[above ] at (0.5,0) {$NP$};
           \node[above] at (0.5,-.5) {$\mbox{John}$};
        \end{scope}
        \node at(5.5,-1) {$\Longrightarrow$};
        \begin{scope}[shift={(8,0)}]

        \begin{scope}
         \draw[thick,->](2.75,-1.5) to  [out=90,in=180] (3.25,-.75) to  [out=0,in=-90] (3.5,0);

         \draw [fill] (2.75,-1.5) circle [radius=0.05];
        \draw[thick,<-](2.25,-1.5) to  [out=90,in=0] (1.75,-0.75) to  [out=180,in=-90] (1.5,0);



        \node[above] at (1,-.6) {$\mbox{``loves''}$};
        \node[above] at (4.2,-.6) {$\mbox{madly}$};
          \node[below] at (2.45,-1.5) {$NP$};
          \node[above] at (3.3,0) {$S$};

        \draw[thick,-](2,0) to  [out=-90,in=180] (2.5,-0.75) to  [out=0,in=-90] (3,0);
        \draw [fill] (3,0) circle [radius=0.05];
        \end{scope}

        \begin{scope}[shift={(-1.5,0)}]
         \draw[thick,-](0,0) to  [out=-90,in=180] (0.5,-1) to  [out=0,in=-90] (1,0);

           \node[above] at (0.5,-.5) {$\mbox{``John''}$};
        \end{scope}

        \draw[thick,-](-1.5,0) to [out=90,in=180] (.25,1) to [out=0,in=90](2,0);
        \draw[thick,-](-.5,0)to [out=90,in=180] (.25,.5) to [out=0,in=90](1.5,0);

        \end{scope}
            \end{tikzpicture}
}
    \\
\scalebox{.75}[1.]
{
    \begin{tikzpicture}

    \begin{scope}[shift={(0,.75)}]
         \node at (0,-1.25) {$=$};
    \begin{scope}
             \draw[thick,->](2.75,-1.5)-- (2.75,0);

             \draw [fill] (2.75,-1.5) circle [radius=0.05];
            \draw[thick,<-](2.25,-1.5) -- (2.25,0);
            \draw [fill] (2.25,0) circle [radius=0.05];

            \node[above] at (1.3,-.6) {$\mbox{John}~\mbox{loves}$};
            \node[above] at (3.35,-.6) {$\mbox{madly}$};
              \node[below] at (2.5,-1.5) {$NP$};

              \node[above] at (2.5,0) {$S$};

    \end{scope}
        \end{scope}

    \node at(5.25,-.5) {$\Longrightarrow$};

    \draw[thick,->](6.5,.75) to  [out=-90,in=180] (9,-1.25) to  [out=0,in=-90] (11.5,.75);
    \draw[thick,<-](7,.75) to  [out=-90,in=180] (9,-.75) to  [out=0,in=-90] (11,.75);
    \draw [fill] (6.5,.75) circle [radius=0.05];
    \draw [fill] (11,.75) circle [radius=0.05];
    \node[above right ] at (11,.75) {$S$};
    \node[above right ] at (6.5,.75) {$NP^\bot$};
    \node[above] at (9,-.6) {$\mbox{John}~\mbox{loves}$};
    \node[above] at (9,-1.22) {$\mbox{madly}$};

    \end{tikzpicture}
}
\\
\scalebox{.75}[1.]
{
\begin{tikzpicture}
\node [left]at(0,0){4)};
\begin{scope}[shift={(-6.5,0)}]
\draw[thick,->](6.5,.75) to  [out=-90,in=180] (9,-1.25) to  [out=0,in=-90] (11.5,.75);
\draw[thick,<-](7,.75) to  [out=-90,in=180] (9,-.75) to  [out=0,in=-90] (11,.75);
\draw [fill] (6.5,.75) circle [radius=0.05];
\draw [fill] (11,.75) circle [radius=0.05];
\node[above right ] at (11,.75) {$S$};
\node[above right ] at (6.5,.75) {$NP^\bot$};
\node[above] at (9,-.6) {$\mbox{John}~\mbox{loves}$};
\node[above] at (9,-1.22) {$\mbox{madly}$};
\end{scope}
\begin{scope}[xscale=1,shift={(6,.5)}]
\draw[thick,->](0,0) to  [out=-90,in=180] (2.5,-1.5) to  [out=0,in=-90] (4.5,0);
\draw [fill] (0,0) circle [radius=0.05];
\node[above] at (0.3,0) {$S^\bot$};

 \draw[thick,->](1.5,0) to  [out=-90,in=180] (1.75,-.75) to  [out=0,in=-90] (2,0);
 \draw [fill] (1.5,0) circle [radius=0.05];
\draw[thick,<-](0.5,0) to  [out=-90,in=180] (1.75,-1.25) to  [out=0,in=-90] (3,0);
\draw [fill] (3,0) circle [radius=0.05];
\draw[thick,<-](3.5,0) to  [out=-90,in=180] (3.75,-0.75) to  [out=0,in=-90] (4.,0);
\draw [fill] (4.,0) circle [radius=0.05];
\node[above] at (1.75,-1.25) {$\mbox{whom}$};
  \node[above] at (1.8,0) {$NP$};
\node[above] at (3.3,0) {$NP^\bot$};
\node[above ] at (4.25,0) {$NP$};
\end{scope}

\node at (11,-.25) {$\Longrightarrow$};
\end{tikzpicture}
}
\\
\scalebox{.75}[1.]
{
\begin{tikzpicture}

\begin{scope}[shift={(-6.5,-.75)}]
\draw[thick,-](6.5,.75) to  [out=-90,in=180] (9,-1.25) to  [out=0,in=-90] (11.5,.75);
\draw[thick,-](7,.75) to  [out=-90,in=180] (9,-.75) to  [out=0,in=-90] (11,.75);
\node[above] at (9,-.6) {$\mbox{``John}~\mbox{loves''}$};
\node[above] at (9,-1.22) {$\mbox{``madly''}$};
\end{scope}
\begin{scope}[xscale=1,shift={(6,0)}]
\draw[thick,->](0,0) to  [out=-90,in=180] (2.5,-1.5) to  [out=0,in=-90] (4.5,0);

 \draw[thick,-](1.5,0) to  [out=-90,in=180] (1.75,-.75) to  [out=0,in=-90] (2,0);
\draw[thick,-](0.5,0) to  [out=-90,in=180] (1.75,-1.25) to  [out=0,in=-90] (3,0);
\draw [fill] (3,0) circle [radius=0.05];
\draw[thick,<-](3.5,0) to  [out=-90,in=180] (3.75,-0.75) to  [out=0,in=-90] (4.,0);
\draw [fill] (4.,0) circle [radius=0.05];
\node[above] at (1.75,-1.25) {$\mbox{``whom''}$};
\node[above] at (3.3,0) {$NP^\bot$};
\node[above ] at (4.25,0) {$NP$};
\end{scope}

\draw[thick,-](4.5,0) to [out=90,in=180] (5.5,1) to [out=0,in=90](6.5,0);
\draw[thick,-](5,0)to [out=90,in=180] (5.5,.5) to [out=0,in=90](6,0);

\draw[thick,-](0,0) to [out=90,in=180] (4,2) to [out=0,in=90](8,0);
\draw[thick,-](.5,0)to [out=90,in=180] (3.5,1.5) to [out=0,in=90](7.5,0);

\node at (11,-.25) {$=$};
\end{tikzpicture}
}
\\
\scalebox{.75}[1.]
{
\begin{tikzpicture}
\begin{scope}[shift={(8.5,0)}]

    \draw[thick,->](0,0) to  [out=-90,in=180] (5,-2) to  [out=0,in=-90] (10,0);
    \draw[thick,<-](0.5,0) to  [out=-90,in=180] (5,-1.5) to  [out=0,in=-90] (9.5,0);
    \draw [fill] (0,0) circle [radius=0.05];
    \draw [fill] (9.5,0) circle [radius=0.05];
    \node[above right ] at (9.5,0) {$NP$};
    \node[above right ] at (0,0) {$NP^\bot$};
    \node[above] at (5,-2) {$\mbox{whom John loves madly}$};

\end{scope}

\end{tikzpicture}

}
\caption{Cowordism representation, steps 2)-4)}
\label{ACG_ex_picture_cont}
\end{subfigure}
\caption{ACG representation example (continued)}
\end{figure}

\section{Linear logic grammars}
Recall that, given a set $N$ of  {\it positive literals} or {\it atoms},
the set  $Fm=Fm(N)$ of  {\it multiplicative linear logic} ({\bf MLL}) formulas over $N$ is defined by the grammar
$Lit::=N|N^\bot$, $Fm::=Lit|Fm\otimes Fm|Fm\parr Fm$.

Connectives $\otimes$ and $\parr$ are called respectively {\it tensor} (also {\it times}) and {\it cotensor} (also {\it par}).

{\it Linear negation} $(.)^\bot$ is not a connective, but is {\it definable}
by induction as
$(P^\bot)^\bot=P$, for $P\in N$, and
$(A\otimes B)^\bot=B^\bot\parr A^\bot$, $(A\parr B)^\bot=B^\bot\otimes A^\bot$.

Note  that, somewhat non-traditionally, we follow the convention that negation flips tensor/cotensor factors,  typical for {\it noncommutative} systems. This does not change the logic (the formulas $A\otimes B$ and $B\otimes A$ are provably equivalent), but is more consistent with the intended interpretation in the category of cowordisms.

\begin{figure}
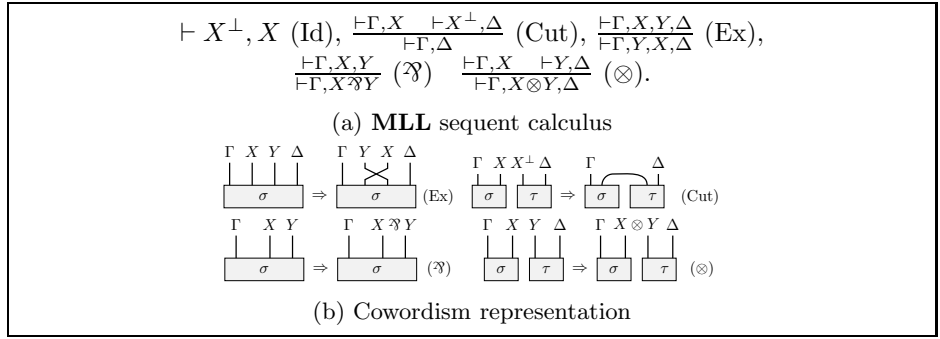

\begin{subfigure}{1\textwidth}
\centering
$\vdash X^\bot,X~(\mbox{Id})$,
$\frac{\vdash \Gamma,X\quad\vdash
X^\bot,\Delta}{\vdash\Gamma,\Delta} ~(\mbox{Cut})$,
$\frac{\vdash
\Gamma,X,Y,\Delta}{\vdash
\Gamma,Y,X,\Delta}
~(\mbox{Ex})$,\\
$\frac{\vdash \Gamma,X,Y}
{\vdash\Gamma,X\parr Y}~
(\parr)\quad\frac{\vdash\Gamma, X \quad\vdash
Y,\Delta}{\vdash\Gamma,X\otimes Y,\Delta}~{ }
(\otimes)$.
\caption{{\bf MLL} sequent calculus}
\label{MLL}
\end{subfigure}
\begin{subfigure}{1\textwidth}
\centering
 \scalebox{.6}
  {
 \tikz[scale=.5]{
  \begin{scope}[shift={(21.5,0)}]
 \begin{scope}[shift={(1,0)}]

 \draw[draw=black,fill=gray!10](-5,0)rectangle(-3.5,1);\node at(-4.25,.5){$\sigma$};
  \draw[draw=black,fill=gray!10](-3,0)rectangle(-1.5,1);\node at(-2.25,.5){$\tau$};
 \draw[thick,-](-4.75,1)--(-4.75,1.5);

 \node[above] at(-4.75,1.5) {$\Gamma$};
 \draw[thick,-](-3.75,1)--(-3.75,1.5);
 \node[above] at(-3.75,1.5) {$X$};

 \draw[thick,-](-1.75,1)--(-1.75,1.5);

 \node[above] at(-1.75,1.5) {$\Delta$};
 \draw[thick,-](-2.75,1)--(-2.75,1.5);
 \node[above] at(-2.75,1.5) {$X^\bot$};
 \node at (-.75,.5) {$\Rightarrow$};
 \end{scope}

\begin{scope}[shift={(-4,0)}]
     \draw[draw=black,fill=gray!10](5,0)rectangle(6.5,1);\node at(5.75,.5){$\sigma$};
 \draw[thick,-](5.25,1)--(5.25,1.5);

 \node[above] at(5.25,1.5) {$\Gamma$};
 \draw[thick,-](5.75,1)to[out=90,in=180](6.25,1.5)
 to[out=0,in=90](7.75,1);

         \draw[draw=black,fill=gray!10](7.,0)rectangle(8.5,1);\node at(8,.5){$\tau$};

 \draw[thick,-](8.25,1)--(8.25,1.5);

 \node[above] at(8.25,1.5) {$\Delta$};
 \node at (10.1,.5) {$(\mbox{Cut})$};
 \end{scope}
 \end{scope}
 \begin{scope}[shift={(11.5,0)}]
 \draw[draw=black,fill=gray!10](-5,0)rectangle(-1.5,1);\node at(-3.25,.5){$\sigma$};
 \draw[thick,-](-4.75,1)--(-4.75,2);

 \node[above] at(-4.75,2.) {$\Gamma$};
 \draw[thick,-](-3.75,1)--(-3.75,2);
 \node[above] at(-3.75,2) {$X$};

 \draw[thick,-](-1.75,1)--(-1.75,2);

 \node[above] at(-1.75,2) {$\Delta$};
 \draw[thick,-](-2.75,1)--(-2.75,2);
 \node[above] at(-2.75,2) {$Y$};
 \node at (-.75,.5) {$\Rightarrow$};

         \draw[draw=black,fill=gray!10](0,0)rectangle(3.5,1);\node at(1.75,.5){$\sigma$};
 \draw[thick,-](.25,1)--(.25,2);

 \node[above] at(.25,2.) {$\Gamma$};
 \draw[thick,-](1.25,1)--(1.25,1.25)--(2.25,1.75)--(2.25,2);
 \node[above] at(2.25,2) {$X$};

 \draw[thick,-](3.25,1)--(3.25,2);

 \node[above] at(3.25,2) {$\Delta$};
 \draw[thick,-](2.25,1)--(2.25,1.25)--(1.25,1.75)--(1.25,2);
 \node[above] at(1.25,2) {$Y$};
 \node at (4.5,.5) {$(\mbox{Ex})$};
 \end{scope}
  }
  }\\

  \scalebox{.6}
  {
  \tikz[scale=.5]{
 \begin{scope}
 \draw[draw=black,fill=gray!10](-5,0)rectangle(-1.5,1);\node at(-3.25,.5){$\sigma$};
 \draw[thick,-](-4.5,1)--(-4.5,2);

 \node[above] at(-4.5,2.) {$\Gamma$};
 \draw[thick,-](-3.,1)--(-3.,2);
 \node[above] at(-3.,2) {$X$};

 \draw[thick,-](-2.,1)--(-2.,2);

 \node[above] at(-2,2) {$Y$};
 \node at (-.75,.5) {$\Rightarrow$};

         \draw[draw=black,fill=gray!10](0,0)rectangle(3.5,1);\node at(1.75,.5){$\sigma$};
 \draw[thick,-](.5,1)--(.5,2);

 \node[above] at(.5,2.) {$\Gamma$};
 \draw[thick,-](2,1)--(2,2);
 \node[above] at(2.5,2) {$X\parr Y$};

 \draw[thick,-](3.,1)--(3.,2);

 \node at (4.5,.5) {$(\parr)$};
 \end{scope}
\begin{scope}[shift={(11.5,0)}]
 \draw[draw=black,fill=gray!10](-5,0)rectangle(-3.5,1);\node at(-4.25,.5){$\sigma$};
  \draw[draw=black,fill=gray!10](-3,0)rectangle(-1.5,1);\node at(-2.25,.5){$\tau$};
 \draw[thick,-](-4.85,1)--(-4.85,2);

 \node[above] at(-4.85,2.) {$\Gamma$};
 \draw[thick,-](-3.75,1)--(-3.75,2);
 \node[above] at(-3.75,2) {$X$};

 \draw[thick,-](-1.65,1)--(-1.65,2);

 \node[above] at(-1.65,2) {$\Delta$};
 \draw[thick,-](-2.75,1)--(-2.75,2);
 \node[above] at(-2.75,2) {$Y$};
 \node at (-.75,.5) {$\Rightarrow$};

         \draw[draw=black,fill=gray!10](0,0)rectangle(1.5,1);\node at(.75,.5){$\sigma$};
 \draw[thick,-](.15,1)--(.15,2);

 \node[above] at(.15,2.) {$\Gamma$};
 \draw[thick,-](1.25,1)--(1.25,2);

         \draw[draw=black,fill=gray!10](2.,0)rectangle(3.5,1);\node at(3,.5){$\tau$};

 \draw[thick,-](3.35,1)--(3.35,2);

 \node[above] at(3.35,2) {$\Delta$};
 \draw[thick,-](2.25,1)--(2.25,2);
 \node[above] at(1.75,2) {$X\otimes Y$};
 \node at (4.65,.5) {$(\otimes)$};
  \end{scope}

 }
 }
\caption{Cowordism representation}
\label{MLL_interpretation}
\end{subfigure}
\caption{Cowordism representation of {\bf MLL}}
\end{figure}

An  ${\bf MLL}$ {\it sequent} (over the alphabet $N$) is a finite sequence of ${\bf MLL}$ formulas (over $N$).
The {\it sequent calculus} for ${\bf MLL}$ \cite{Giriard_TCS} is shown in Figure \ref{MLL}.

It is well known \cite{Seely} that semantics of {\bf MLL} proof theory is provided by  {\it $*$-autonomous categories}. Compact categories are a particular (degenerate) case of these, so
 the  category ${\bf Coword}_T$ of cowordisms over an alphabet $T$ allows interpretation of {\bf MLL} (invariant under cut-elimination).

Just as in the case of linear $\lambda$-calculus (and {\bf ILL}), an interpretation $\xi$ consists in assigning to every atom $A\in N$ a boundary $\xi(A)$.
This is extended to all formulas in $Fm(N)$ by
$\xi(A\otimes B)=\xi(A\parr B)=\xi(A)\otimes \xi(B)$ and $\xi(A^\bot)=\xi(A)^\bot$ (note that the extension is well defined).

A sequent $\Gamma=A_1,\ldots A_n$ is interpreted as the cotensor of its formulas:
$\xi(\Gamma)=\xi(A_1\parr\ldots\parr A_n)=\xi(A_1)\otimes\ldots\otimes\xi(A_n)$.
A proof $\sigma$ of the sequent $\vdash \Gamma$ is interpreted as a multiword with boundary
$\xi(\Gamma)$, which we  identify with a
cowordism
$
\xi(\sigma):{\bf 1}\to \xi(\Gamma)$.
Rules for interpreting sequent calculus proofs are  represented in Figure \ref{MLL_interpretation} (the symbol $\xi$ omitted and picture rotated counterclockwise with outgoing boundaries on the top, as before).

Given an interpretation $\xi$ of $Fm(N)$ in the category ${\bf Coword_T}$, we say
 that a {\it cowordism typing judgement}  (over $N$ and $T$) is an expression of the form $\cfrac{\sigma}{\vdash \Gamma}$, where $\Gamma$ is an ${\bf MLL}$ sequent (over  $N$), and $\sigma:{\bf 1}\to \xi(\Gamma)$ is a cowordism (over $T$).

A  {\it linear logic grammar} {\bf LLG} $G$ is a tuple  $G=(N,\xi,T,Lex,S)$, where
$N$, $\xi$, $T$  are as above, while
     $Lex$, the {\it lexicon}, is a finite set of cowordism typing judgements over $N$ and $T$, called  {\it axioms}, and
   $S\in N$, the {\it initial type}, is a positive literal with $|\xi(S)|=2$ and $(\xi(S))_l$ a singleton.

We say  that
the cowordism typing judgement $\cfrac{\sigma}{\vdash \Gamma}$ is {\it derivable in $G$}, or that
$G$ {\it generates  cowordism $\sigma$  of type $\Gamma$}
if there exists a derivation of
$\vdash \Gamma$ from axioms of $G$ whose interpretation is $\sigma$.

 Any regular cowordism of the initial type $S$ generated by $G$ is an edge-labeled graph containing  a single edge labeled with a word over $T$. Thus the set of type $S$ regular cowordisms can be identified with a set of words in $T^*$.
The {\it language $L(G)$ generated by $G$} is the set of words labeling type $S$ regular cowordisms generated by $G$.

\begin{thm}
A   language  generated by a string ACG   is also generated by an LLG.
\end{thm}
{\bf Proof} Given a string ACG $G=(\Sigma,T, \phi,S)$ over the set $N$ of atomic types and the terminal alphabet $T$, we identify types of $\Sigma$  with a subset of the set $Fm(N)$ of {\bf MLL} formulas  using the translation $A\multimap B=A^\bot \parr B$.

Then the cowordism representation $\xi$ of $G$  gives us  an interpretation  of $Fm(N)$ in ${\bf Coword_T}$. 
Taking as the lexicon
$Lex$ the set of all  cowordism typing judgements $\cfrac{\xi(c)}{\vdash A}$, where $\vdash c:A$ is an axiom of $\Sigma$,
we obtain the LLG
$G'=(N,\xi,T,Lex,S)$.

By induction on derivations it can be shown that for  any cowordism $\sigma$ of the form $\sigma:\xi(A_1)\otimes\ldots\otimes\xi(A_n)\to\xi(A)$, where  $A_1,\ldots,A_n,A$ are in $Tp(N)$, the cowordism typing judgement
$\cfrac{\sigma}{\vdash A_n^\bot,\ldots, A_1^\bot,A}$ is derivable in $G'$ iff $\sigma$ is the cowordism representation of some typing judgement $A_1,\ldots,A_n\vdash A$ derivable in $\Sigma$. (Essentially, this repeats the proof that {\bf ILL} is a conservative fragment of {\bf MLL}.) The statement follows. $\Box$

It seems reasonable to ask whether the converse is true. We would expect that the answer is yes, and the formalism of LLG does not add extra expressivity.

\section{LLG and multiple context-free grammars}
We discuss relations between LLG and {\it multiple context-free grammars}.

   Assume that we are given   a finite  alphabet $N$ of nonzero arity predicate symbols  called {\it nonterminal symbols} 
    and
     a finite alphabet $T$ of {\it terminal symbols}. 

  {\it Production} is a sequent
    of the form
  \begin{equation}\label{production}
   B_1(x^1_1,\ldots,x^1_{k_1}),\ldots,B_n(x^n_1,\ldots,x^n_{k_n})\vdash A(s_1,\ldots,s_k),
   \end{equation}
    where $A,B_1,\ldots, B_n\in N$ have arities $k,k_1,\ldots,k_n$ respectively,
  $\{x^j_i\}$ are pairwise distinct variables not from $T$, and
  $s_1,\ldots, s_k$ are  words built of terminal symbols and $\{x_i^j\}$,
  so that each of the variables $x^j_i$ occurs exactly once in exactly one of  $s_1,\ldots s_k$ (here $n$ may be zero).

  A {\it multiple context-free grammar (MCFG)}  \cite{Seki} $G$ is a tuple $G=(N,T,S,P)$ where $N,T$ are as above, $P$ is a finite set of productions, and
  $S\in N$, the {\it initial symbol},  is unary.

The set of {\it  predicate formulas derivable in $G$} is defined  by the following induction.

 Formula $A(t_1,\ldots,t_k)$ is derivable, if there is a production
 of the form {(\ref{production})} in $P$, such that
   $B_1(s^1_1,\ldots,s^1_{k_1}),\ldots,B_n(s^n_1,\ldots,s^n_{k_n})$
are derivable, and
   $t_m$ is  the result of substituting the word $s^j_i$ for every variable $x^j_i$ in $s_m$,    $m=1,\ldots,k$.
(The case $n=0$ is the base of induction.)

The {\it language generated } by an MCFG $G$ is the set of words $w$ for which $ S(w)$ is derivable.

 It is well known that  any MCFG translates to a string ACG \cite{Salvati}, hence to an LLG as well.

\begin{figure}
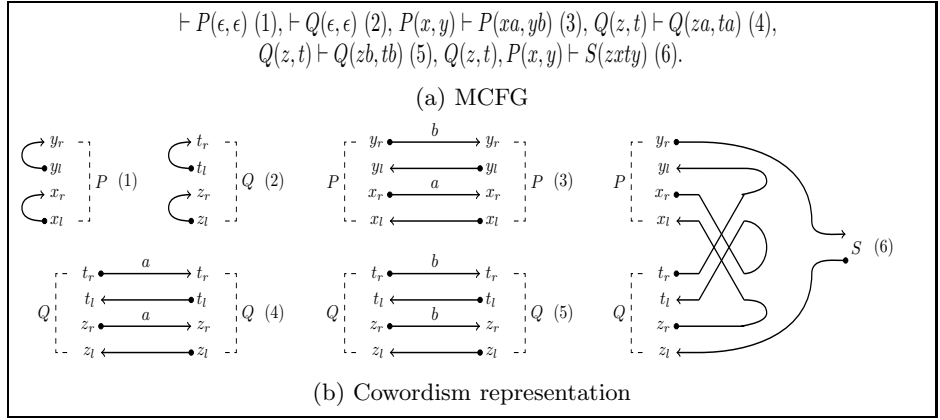

\begin{subfigure}{1\textwidth}
\centering
\scalebox{.7}[1]
{
\text{
$\vdash P(\epsilon,\epsilon)~(1)$, $\vdash Q(\epsilon,\epsilon)~(2)$,  $P(x,y)\vdash P(xa,yb)~(3)$,
$Q(z,t)\vdash Q(za,ta)~(4)$,}}\\
\scalebox{.7}[1]
{
\text{
  $Q(z,t)\vdash Q(zb,tb)~(5)$,
$Q(z,t),P(x,y)\vdash S(zxty)~(6)$.
}
}
\caption{MCFG}
\label{MCFG}
\end{subfigure}
\begin{subfigure}{1\textwidth}
\scalebox{.6}[.7]
{
 \tikz[scale=1]{
 \node[right] at(-2.5,.75) {$(1)$};


\node[right] at(-4,0) {$x_l$};
 \node[right] at(-4,.5) {$x_r$};
 \node[right] at(-4,1) {$y_l$};
\node[right] at(-4,1.5) {$y_r$};

\draw[dashed,-](-3.25,1.5)--(-3,1.5)--(-3,0)--(-3.25,0);
\node[right]at (-3,.75){$P$};

\draw [fill] (-4,0) circle [radius=0.05];
\draw [fill] (-4,1) circle [radius=0.05];

\draw[thick,->](-4,0)
to[out=180, in=-90] (-4.5,.25)
to[out=90, in=180] (-4,.5);

\draw[thick,->](-4,1)
to[out=180, in=-90] (-4.5,1.25)
to[out=90, in=180] (-4,1.5);

\begin{scope}[shift={(-1,0)}]
\node[right] at(1.75,.75) {$(2)$};

\draw[dashed,-](1,1.5)--(1.25,1.5)--(1.25,0)--(1,0);
\node[right]at (1.25,.75){$Q$};


\node[right] at(.25,0) {$z_l$};
 \node[right] at(.25,.5) {$z_r$};
 \node[right] at(.25,1) {$t_l$};
\node[right] at(.25,1.5) {$t_r$};

\draw [fill] (.25,0) circle [radius=0.05];
\draw [fill] (.25,1) circle [radius=0.05];

\draw[thick,->](.25,0)
to[out=180, in=-90] (-.25,.25)
to[out=90, in=180] (.25,.5);

\draw[thick,->](.25,1)
to[out=180, in=-90] (-.25,1.25)
to[out=90, in=180] (.25,1.5);
\end{scope}
\begin{scope}[shift={(-1.85,0)}]
\node[right] at(9,.75) {$(3)$};

\draw[dashed,-](8.25,1.5)--(8.5,1.5)--(8.5,0)--(8.25,0);
\node[right]at (8.5,.75){$P$};

\draw[dashed,-](4.75,1.5)--(4.5,1.5)--(4.5,0)--(4.75,0);
\node[left]at (4.5,.75){$P$};

\node[left] at(5.5,0) {$x_l$};
 \node[left] at(5.5,.5) {$x_r$};
 \node[left] at(5.5,1) {$y_l$};
\node[left] at(5.5,1.5) {$y_r$};

\node[right] at(7.5,0) {$x_l$};
 \node[right] at(7.5,.5) {$x_r$};
 \node[right] at(7.5,1) {$y_l$};
\node[right] at(7.5,1.5) {$y_r$};

\draw [fill] (5.5,.5) circle [radius=0.05];
\draw [fill] (5.5,1.5) circle [radius=0.05];
\draw [fill] (7.5,0) circle [radius=0.05];
\draw [fill] (7.5,1) circle [radius=0.05];

\node[above] at(6.5,1.5) {$b$};
\node[above] at(6.5,0.5) {$a$};

\draw[thick,<-](5.5,0)--(7.5,0);
\draw[thick,->](5.5,.5)--(7.5,.5);
\draw[thick,<-](5.5,1)--(7.5,1);
\draw[thick,->](5.5,1.5)--(7.5,1.5);
\end{scope}

\begin{scope}[shift={(-2.75,-2.5)}]
    \node[right] at(3.5,.75) {$(4)$};

    \draw[dashed,-](-.75,1.5)--(-1,1.5)--(-1,0)--(-.75,0);
    \node[left]at (-1,.75){$Q$};

    \draw[dashed,-](2.75,1.5)--(3,1.5)--(3,0)--(2.75,0);
    \node[right]at (3,.75){$Q$};

    \node[left] at(0,0) {$z_l$};
     \node[left] at(0,.5) {$z_r$};
     \node[left] at(0,1) {$t_l$};
    \node[left] at(0,1.5) {$t_r$};

    \node[right] at(2,0) {$z_l$};
     \node[right] at(2,.5) {$z_r$};
     \node[right] at(2,1) {$t_l$};
    \node[right] at(2,1.5) {$t_r$};

    \draw [fill] (0,.5) circle [radius=0.05];
    \draw [fill] (0,1.5) circle [radius=0.05];
    \draw [fill] (2,0) circle [radius=0.05];
    \draw [fill] (2,1) circle [radius=0.05];

    \node[above] at(1,1.5) {$a$};
    \node[above] at(1,0.5) {$a$};

    \draw[thick,<-](0,0)--(2,0);
    \draw[thick,->](0,.5)--(2,.5);
    \draw[thick,<-](0,1)--(2,1);
    \draw[thick,->](0,1.5)--(2,1.5);
\begin{scope}[shift={(-.85,0)}]
    \node[right] at(10.75,.75) {$(5)$};

    \draw[dashed,-](6.5,1.5)--(6.25,1.5)--(6.25,0)--(6.5,0);
    \node[left]at (6.25,.75){$Q$};

    \draw[dashed,-](10,1.5)--(10.25,1.5)--(10.25,0)--(10,0);
    \node[right]at (10.25,.75){$Q$};

    \node[left] at(7.25,0) {$z_l$};
     \node[left] at(7.25,.5) {$z_r$};
     \node[left] at(7.25,1) {$t_l$};
    \node[left] at(7.25,1.5) {$t_r$};

    \node[right] at(9.25,0) {$z_l$};
     \node[right] at(9.25,.5) {$z_r$};
     \node[right] at(9.25,1) {$t_l$};
    \node[right] at(9.25,1.5) {$t_r$};

    \draw [fill] (7.25,.5) circle [radius=0.05];
    \draw [fill] (7.25,1.5) circle [radius=0.05];
    \draw [fill] (9.25,0) circle [radius=0.05];
    \draw [fill] (9.25,1) circle [radius=0.05];

    \node[above] at(8.25,1.5) {$b$};
    \node[above] at(8.25,0.5) {$b$};

    \draw[thick,<-](7.25,0)--(9.25,0);
    \draw[thick,->](7.25,.5)--(9.25,.5);
    \draw[thick,<-](7.25,1)--(9.25,1);
    \draw[thick,->](7.25,1.5)--(9.25,1.5);
\end{scope}
\end{scope}

\begin{scope}[shift={(10,-2.5)}]
\node[right] at(4.25,2) {$(6)$};

\node[left] at(0,0) {$z_l$};
 \node[left] at(0,.5) {$z_r$};
 \node[left] at(0,1) {$t_l$};
\node[left] at(0,1.5) {$t_r$};
 \node[left] at(0,2.5) {$x_l$};
\node[left] at(0,3) {$x_r$};
 \node[left] at(0,3.5) {$y_l$};
\node[left] at(0,4) {$y_r$};

\draw[thick,<-](0,0)--(.5,0);
\draw[thick,-](0,.5)--(.5,.5);
\draw[thick,<-](0,1)--(.5,1);
\draw[thick,-](0,1.5)--(.5,1.5);
\draw[thick,<-](0,2.5)--(.5,2.5);
\draw[thick,-](0,3)--(.5,3);
\draw[thick,<-](0,3.5)--(.5,3.5);
\draw[thick,-](0,4)--(.5,4);

\draw[dashed,-](-.75,4)--(-1,4)--(-1,2.5)--(-.75,2.5);

\draw[dashed,-](-.75,1.5)--(-1,1.5)--(-1,0)--(-.75,0);

\node[left] at(-1,.75) {$Q$};
\node[left] at(-1,3.25) {$P$};

\node[right] at(3.75,2) {$S$};

\draw[thick,->](.5,4)
to[out=0,in=90](3.,2.5)
to[out=-90,in=180](3.75,2.25)
;

\draw[thick,-](.5,0)
to[out=0,in=-90](3.,1.5)
to[out=90,in=180](3.75,1.75)
;

\draw [fill] (3.75,1.75) circle [radius=0.05];

\draw[thick,-](0.5,3)--(1.5,1.5);
\draw[thick,-](0.5,1.5)--(1.5,3);
\draw[thick,-](0.5,2.5)--(1.5,1.);
\draw[thick,-](0.5,1.)--(1.5,2.5);


\draw[thick,-](1.5,1.5)
to[out=0,in=-90](2.,2)
to[out=90,in=0](1.5,2.5)
;

\draw[thick,-](1.5,3)
to[out=180,in=-90](2,3.25)
to[out=90,in=0](1.5,3.5)
--(.5,3.5)
;

\draw[thick,-](0.5,.5)--(1.5,.5)
to[out=180,in=-90](2,.75)
to[out=90,in=0](1.5,1)
;

 \draw [fill] (0,.5) circle [radius=0.05];
\draw [fill] (0,1.5) circle [radius=0.05];
\draw [fill] (0,3) circle [radius=0.05];
\draw [fill] (0,4) circle [radius=0.05];

\end{scope}

 }
 }
 \caption{Cowordism representation}
 \label{MCFG_cowordisms}
 \end{subfigure}
 \caption{Cowordism representation of an MCFG}
 \label{MCFG_example}
 \end{figure}

A concrete example of a {\it cowordism representation} for an MCFG is given in Figure \ref{MCFG_example}.
Here we have the terminal alphabet $T=\{a,b\}$, and nonterminal symbols  $P$, $Q$ and $S$  of arities 2, 2 and 1 respectively. The MCFG is defined by the six productions in Figure \ref{MCFG}.
It is easy to see that the above generates the language $\{wa^nwb^n|w\in T^*, n\geq 0\}$.

Six cowordisms representing the productions are shown in Figure \ref{MCFG_cowordisms} (for better readability, we label vertices with corresponding variables,  the subscripts $l,r$ denoting  left and right endpoints respectively).
In order to turn these into axioms for an LLG,  we  have to get rid of the incoming wires.  We make all wires outgoing using the bijection $Hom(X,Y)\cong Hom({\bf 1}, X^\bot\otimes Y)$ (which is a particular case of (\ref{compact structure}), whose geometric meaning is shown in Figure \ref{compactness_picture}).

\begin{thm}
 A language is generated  by an MCFG iff it can be generated by an LLG $G$ with {\it $\otimes$-free} lexicon.
\end{thm}
{\bf Proof}
Translation from MCFG to a ({$\otimes$}-free) LLG is easy, an example has just been shown. Let us prove the other direction.

For a boundary $X$ and a regular multiword $M$ with boundary $X$, we define the {\it pattern} $pat(M)$ of $M$ as the  graph obtained by erasing from $M$ all letters. The set $Patt(X)$ of all graphs obtained in this way as $M$ varies is the set of {\it possible patterns} of $X$. Note that $Patt(X)$ is finite (maybe empty).

Now, for any $\pi\in Patt(X)$ choose an enumeration of edges in $\pi$ and introduce a $k$-ary predicate symbol $X^\pi$, where $k$ is the number of edges in $\pi$ (obviously, $k$ is the same for all possible patterns of  $X$). Then any regular multiword $M$ with boundary $X$ can be unambiguously represented as the predicate formula $X^\pi(w_1,\ldots,w_k)$, where $\pi=pat(M)$, and $w_i$ is the word labeling  the $i$-th edge of $\pi$ in $M$, $i=1,\ldots,k$.

In a similar way, any cowordism $\sigma:X_1\otimes\ldots\otimes X_n\to X$ can be encoded into a finite set of productions. (The above described representation of a multiword is a particular case when $n=0$).

Fix possible patterns  $\pi_1,\ldots,\pi_{n}$ of
$X_1,\ldots X_{n}$ respectively. There exists at most one possible pattern $\pi$ of $X$ such that, whenever
$pat(M_i)=\pi_i$, $i=1,\ldots,n$,  it
holds that
$pat(\sigma\circ(M_1\otimes\ldots\otimes M_{n}))=\pi$.
If such a $\pi$ does not exist, then  the chosen combination of patterns composed with $\sigma$ does not produce a regular multiword and is irrelevant for us.

Otherwise choose fresh variables $x^i_j$, $j=1,\ldots,k_i$,  where $k_i$ is the number of edges in $\pi_i$,
$i=1,\ldots,n$. Let $M_i$ be the multiword obtained from $\pi_i$ by labeling the $j$-th edge with $x^i_j$. Let
$M=\sigma\circ(M_1\otimes\ldots\otimes M_{n})$. It is a multiword with $pat(M)=\pi$. Let $s_j$ be the word labeling the $j$-th edge of $M$, $j=1,\ldots,k$, where $k$ is the number of edges in $\pi$.
The interaction of $\sigma$ with the chosen combination of patterns is represented as the production
$X^{\pi_1}_1(x^1_1,\ldots,x^1_{k_1}),\ldots, X^{\pi_n}_n(x^n_1,\ldots,x^n_{k_n})\vdash X^\pi(s_1,\ldots,s_k)$.

Let $Prod(\sigma)$ be the set of all productions obtained in this way from $\sigma$ by varying possible patterns of $X_1,\ldots,X_n$. Again, note that $Prod(\sigma)$ is finite.

Now, let $G=(N,\xi,T,S,Lex)$ be a $\otimes$-free LLG. The symbol $\xi$ will be omitted in what follows.

  We know that a sequent $\vdash\Gamma,A\parr B$ is derivable in {\bf MLL} iff  $\vdash\Gamma,A,B$ is. And since  axioms of $G$ do not use any connective other than $\parr$, it follows that $G$ is equivalent to a grammar that does not use any logical connective at all.  By cut-elimination, any derivation of the sequent $\vdash S$ from  axioms of $G$ is equivalent to a derivation not using any logical rule either, i.e.  to a one  using only the Cut rule.

 We construct an equivalent MCFG $G'=(N',T,S',P)$, by taking the set of nonterminal symbols
 $N'=\{A^\pi|~A\in N\cup N^\bot, \pi\in Patt(A)\}$, and writing for each axiom $\alpha\in Lex$ of the form
$\cfrac{\sigma}{\vdash A_1,\ldots,A_n}$,  where $A_1,\ldots,A_n$ are literals, all productions representing
 cowordisms
$$
\sigma_i:A_{i+1}^\bot\otimes\ldots\otimes A_{n}^\bot\otimes A_1^\bot\otimes\ldots\otimes A_{i-1}^\bot \to A_i,\quad i=1,\ldots,n,$$
obtained from $\sigma$ using correspondence (\ref{compact structure}) and symmetry transformations.

We put $Prod(\alpha)=\bigcup\limits_iProd(\sigma_i)$, and then
$P=\bigcup\limits_{\alpha\in Lex} Prod(\alpha)$.

As for the initial symbol $S'$ of $G'$, we observe that there is only one possible pattern $s$ for the boundary $S$, and we put $S' =S^s$. An easy induction on derivations shows that $G$ and $G'$ generate the same language. $\Box$

As a corollary we  obtain  the known result that  any second order ACG generates a multiple context-free language \cite{Salvati}. Thus, we gave a new, geometric proof, arguably quite simple and intuitive.

\section{Backpack problem}
  An LLG  of a general form can generate an NP-complete language, just as an ACG (see \cite{YoshinakaKanazawa}).   We give the following, last example as another try to convince the reader that the geometric language of cowordisms is indeed intuitive and convenient for analyzing language generation.

We will  consider the backpack problem in the form of the {\it subset sum problem} (SSP):
 given a finite sequence $s$ of integers, determine if there is a subsequence $s'\subseteq s$ such that $\sum\limits_{z\in s'}z=0$. It is well known \cite{Martello} that SSP is NP-complete. We will generate by means of an LLG an NP-complete language, essentially representing solutions of SSP.

We  represent integers as words in the alphabet $\{+,-\}$, we call them {\it numerals}. An integer $z$ is represented (non-uniquely) as a word for which the difference of $+$ and $-$ occurrences equals $z$.
We say that a numeral is {\it irreducible}, if it consists only of pluses or only of minuses.

We represent finite sequences of integers as words in the alphabet $T=\{+,-,\bullet\}$, with $\bullet$ interpreted as a separation sign. Thus a word in this alphabet should be read as a list of numerals separated by bullets.
When all numerals in the list are irreducible, we say that the list is irreducible.


Now consider two positive literals $H,S$ and interpret each of them as the boundary $X$ of cardinality 2 with $X_l=\{2\}$.

\begin{figure}
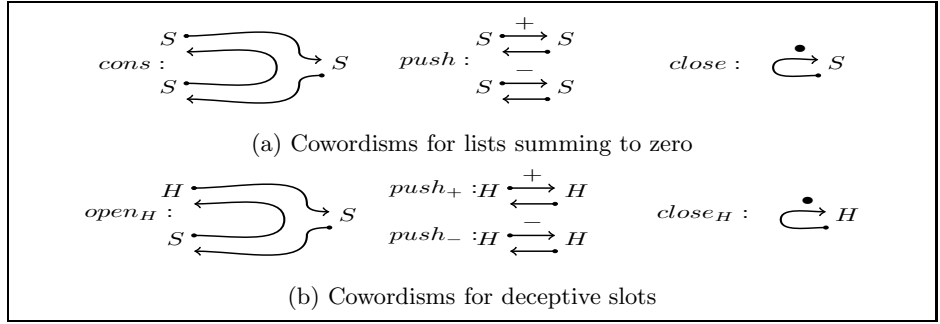

\begin{subfigure}{1\textwidth}
\centering
\scalebox{1}[.7]
{
\tikz[scale=.6]
        {

         \draw[thick,->](0,0) to  [out=0,in=90] (2.5,-0.5) to  [out=-90,in=180] (3,-0.75);
        \draw [fill] (0,0) circle [radius=0.05];

        \draw[thick,<-](0,-0.5) to  [out=0,in=90] (2,-1) to  [out=-90,in=0] (0,-1.5);
        \draw [fill] (-0,-1.5) circle [radius=0.05];
          \draw[thick,<-](0,-2.)to  [out=0,in=-90](2.5,-1.5) to  [out=90,in=-180] (3,-1.25);
          \draw [fill] (3,-1.25) circle [radius=0.05];
          \node[above left] at (0,-0.5) {$S$};
        \node[above left] at (0,-2) {$S$};
        \node[above right] at (3,-1.25) {$S$};
        \node[above left] at(-.25,-1.25) {$cons:$};
        \begin{scope}[shift={(7,0)}]

        \draw [fill] (0,0) circle [radius=0.05];
        \draw[thick,->](0,0) -- (1,0);

        \draw [fill] (1,-0.5) circle [radius=0.05];
        \draw[thick,<-](0,-0.5) -- (1,-0.5);
        \node[above right] at (1,-.5) {$S$};
        \node[above left] at (0,-.5) {$S$};
        \node[above] at (0.5,0) {$+$};

        \draw [fill] (0,-1.5) circle [radius=0.05];
        \draw[thick,->](0,-1.5) -- (1,-1.5);
        \draw [fill] (1,-2) circle [radius=0.05];
        \draw[thick,<-](0,-2.) -- (1,-2.);
        \node[above right] at (1,-2) {$S$};
        \node[above left] at (0,-2) {$S$};
        \node[above] at (0.5,-1.5) {$-$};

        \node[above left] at(-.5,-1.25) {$push:$};
%
        \begin{scope}[shift={(6,-.75)}]
          \node[above] at (0.6,0) {$\bullet$};
        \draw[thick,<-](1,0) to  [out=-180,in=90] (0,-0.25) to  [out=-90,in=-180] (1,-.5);
        \draw [fill] (1,-0.5) circle [radius=0.05];

        \node[above right] at (1,-.5) {$S$};
        \node[above left] at(-.5,-.5) {$close:$};
        \end{scope}
%
        \end{scope}
         }
        }
\caption{Cowordisms for lists summing to zero}
\label{backpack1}
\end{subfigure}
\begin{subfigure}{1\textwidth}
\centering
\scalebox{1}[.7]
{
 \tikz[scale=.6]{
\draw [fill] (0,0) circle [radius=0.05];
 \draw[thick,->](0,0) to  [out=0,in=90] (2.5,-0.5) to  [out=-90,in=180] (3,-0.75);

\draw [fill] (0,-1.5) circle [radius=0.05];
\draw[thick,<-](0,-0.5) to  [out=0,in=90] (2,-1) to  [out=-90,in=0] (0,-1.5);
\draw [fill] (3,-1.25) circle [radius=0.05];
  \draw[thick,<-](0,-2.)to  [out=0,in=-90](2.5,-1.5) to  [out=90,in=-180] (3,-1.25);
  \node[above left] at (0,-0.5) {$H$};
\node[above left] at (0,-2) {$S$};
\node[above right] at (3,-1.25) {$S$};
\node[above left] at(-.25,-1.25) {$open_H:$};
\begin{scope}[shift={(7,0)}]
\begin{scope}[shift={(6,-.75)}]
      \node[above] at (0.6,0) {$\bullet$};
    \draw[thick,<-](1,0) to  [out=-180,in=90] (0,-0.25) to  [out=-90,in=-180] (1,-.5);
    \draw [fill] (1,-.5) circle [radius=0.05];
    \node[above right] at (1,-.5) {$H$};
    \node[above left] at(-.5,-.5) {$close_H:$};
\end{scope}
  \draw [fill] (0,0) circle [radius=0.05];
\draw[thick,->](0,0) -- (1,0);
\draw [fill] (1,-0.5) circle [radius=0.05];
\draw[thick,<-](0,-0.5) -- (1,-0.5);
\node[above right] at (1,-.5) {$H$};
\node[above left] at (0,-.5) {$H$};
\node[above] at (0.5,0) {$+$};
\node[above left] at(-.5,-.5) {$push_+:$};
\begin{scope}[shift={(0,-1.5)}]
  \draw [fill] (0,0) circle [radius=0.05];
\draw[thick,->](0,0) -- (1,0);
\draw [fill] (1,-0.5) circle [radius=0.05];
\draw[thick,<-](0,-0.5) -- (1,-0.5);
\node[above right] at (1,-.5) {$H$};
\node[above left] at (0,-.5) {$H$};

\node[above] at (0.5,0) {$-$};
\node[above left] at(-.5,-.5) {$push_-:$};
\end{scope}
\end{scope}
 }
 }
\caption{Cowordisms for deceptive slots}
\label{backpack2}
\end{subfigure}
\caption{Encoding the backpack problem}
\end{figure}

First we construct a system of cowordisms which, together with symmetry transformations, generates all irreducible lists of integers that sum to zero.
The  three cowordisms
$cons:S\otimes S\to S$,
 $push:S\otimes S\to S\otimes S$, $close:{\bf 1}\to S$
are shown in Figure \ref{backpack1}.
The cowordism $cons$, by iterated compositions with itself, generates lists with arbitrary many empty slots, then $push$ fill the slots with pluses and minuses (always in pairs), and  $close$ closes them. All generated lists will sum to zero, and all irreducible lists summing to zero will be generated.

Now, in order to generate solutions of SSP we need some extra, ``deceptive'' slots, which  contain elements not summing to zero. These slots  will be represented by the boundary $H$.
The corresponding   cowordisms
$open_H:H\otimes S\to S$,
$ push_+:H\to H$, $push_-:H\to H$, $close_H:{\bf 1}\to H$
are shown in Figure \ref{backpack2}.
The cowordism $open_H$ adds deceptive slots to the list, $push_-$ and $push_+$ fill them with arbitrary numerals, and $close_H$ closes them.

Although, pedantically speaking, the above  system is not an LLG according to our definitions, we obtain an LLG by making all wires outgoing, just as in the above discussion of MCFG.
Let us denote the generated language as $L_0$.

It is easy to see that  $L_0$ membership problem is, essentially,  SSP.  More precisely {\it SSP polynomially reduces to $L_0$ membership problem.}
Indeed,  a sequence $s$ of integers is a solution of SSP iff the corresponding irreducible list is in $L_0$.
It follows that $L_0$ is NP-hard.

On the other hand, it is also easy to see that $L_0$ membership problem is itself in NP. Indeed, in order to show that a list $s$ is in $L_0$ it is sufficient to demonstrate a sequence of cowordisms from Figures \ref{backpack1}, \ref{backpack2} generating $s$, and the number of cowordisms in such a sequence clearly is bounded linearly in the size of $s$. Thus $L_0$ is, in fact, NP-complete.

\bibliography{sslavnovbibliography}
\bibliographystyle{alpha}

\end{document}